\documentclass[preprint,preprintnumbers,amsmath,amssymb,floatfix,nofootinbib]{revtex4}
  
\usepackage{epsf}
\usepackage{epsfig}
\usepackage{graphicx}
\begin{document}  

\preprint{BNL-HET-07/}
\title{\vspace*{0.7in}
 SUSY QCD Corrections to Associated Higgs-bottom Quark Production}

\author{S.~Dawson and C.~B.~Jackson}

\affiliation{
Department of Physics, Brookhaven National Laboratory, 
Upton, NY 11973, USA 
\vspace*{.5in}}


\begin{abstract}
The associated production of a Higgs boson with a $b$ quark is a discovery mode for an MSSM Higgs boson at large $\tan\beta$. We compute the SUSY QCD 
corrections from gluino and squark loops 
to this process and combine them with the ${\cal O}(\alpha_s^2)$ NLO QCD 
corrections to obtain reliable predictions for the rate.  Finally,
we compare our results with an effective Lagrangian approximation which 
includes only the low energy corrections from squark and gluino loops to the $b{\overline b}$ 
Higgs vertices. 
\end{abstract}

\maketitle

\newpage

\section{Introduction}

The search for the Higgs boson is among the most important challenges facing
the current generation of colliders.  If there is a single Higgs boson 
with the properties
predicted by the Standard Model, we expect that it will be discovered at either
the Tevatron or the LHC, with the optimal channel for discovery depending strongly
on the Higgs mass.  In the Standard Model, the production of a Higgs
boson in association with $b$ quarks is never important. 
 However, in the minimal supersymmetric model (MSSM), there
are five Higgs bosons--two neutral Higgs bosons, 
$h^0$ and $H^0$, a pseudoscalar, $A^0$,
and two charged Higgs bosons, $H^\pm$-- and the
strategy for discovery is quite different from in the Standard Model.  In 
the MSSM, the couplings of the Higgs bosons to  $b$ quarks can be significantly
enhanced and for a large range of parameter space, Higgs production in association
with $b$ quarks is the most likely discovery channel\cite{Dawson:2005vi,Carena:2007aq,Campbell:2004pu,Dittmaier:2003ej,Carena:1998gk}.  

The production of a Higgs boson in association with a $b$ quark 
has been extensively studied\cite{Dawson:2004sh,Dawson:2005vi,Carena:2007aq,Campbell:2004pu,Dittmaier:2003ej,Carena:1998gk,Dawson:2003kb,Maltoni:2003pn,Maltoni:2005wd,Dicus:1998hs,Campbell:2002zm} and the NLO QCD corrections
are well understood, 
both in the $4$- and $5$- flavor number parton 
schemes\cite{Dawson:2005vi,Campbell:2004pu,Maltoni:2003pn}.  
In the $4$- flavor number scheme, the lowest order processes 
for producing a Higgs boson and a $b$ quark are $gg\rightarrow
 b {\overline b}\phi$ and 
$ q {\overline q}\rightarrow b {\overline b} 
\phi$\cite{Dittmaier:2003ej,Dawson:2004sh,Dicus:1998hs}.
(The neutral
Higgs bosons are generically, $\phi=h^0,H^0,A^0$).
In a $5$- flavor number scheme, the $b$ quark appears as a parton
and potentially large logarithms of the form $\ln({M_\phi\over m_b})$
are absorbed into $b$ quark parton distribution functions 
(PDFs)\cite{Barnett:1987jw,Olness:1987ep}. 
In the $5$- flavor number scheme, the lowest order process for producing
a Higgs boson in association with $b$ quarks is $b{\overline b}
\rightarrow \phi$ when no $b$ quarks are tagged in the final state and 
 $b g\rightarrow b \phi$ when a single $b$ quark 
is tagged\cite{Dicus:1998hs,Dawson:2005vi,Campbell:2004pu,Dawson:2004sh,Dittmaier:2003ej}. 
Contributions from the $gg$  and $q {\overline q} $ initial states
 are thus subleading in the $5-$ flavor number scheme.  
  Although
the $4$- and $5$- flavor number schemes represent different orderings of
perturbation theory, the two schemes have been shown to yield
equivalent numerical results.
The residual renormalization and factorization scale, scheme,
and PDF uncertainties of the NLO calculations result in a $10-20\%$
uncertainty on the prediction of the rate\cite{Dawson:2005vi,Campbell:2004pu}.

When no $b$ quarks are tagged in the final state, the dominant production
mechanism in the $5$-flavor number scheme is $b {\overline b}\rightarrow
\phi$, which has been calculated to NNLO\cite{Harlander:2003ai}.  
When a single $b$ quark is tagged, the rate is lower, but so also is
the background and it is this channel which we focus on here.
Both CDF and D0 have derived limits on $\tan\beta$ from Tevatron data, 
based on the 
search for $p {\overline p}\rightarrow b\phi$ with $\phi\rightarrow
\tau^+\tau^-$ and $\phi\rightarrow b {\overline b}$\cite{Anastassov:2006gc}.
In this paper,  we present the ${\cal O}(\alpha_s^2)$
SUSY QCD (SQCD) corrections from gluino-squark
loops to the $b$- Higgs production processes.
Gluino-squark virtual contributions have
been shown to be significant for the inclusive
$b{\overline b}\rightarrow \phi$
channel\cite{Dittmaier:2006cz} 
and it is of interest to examine the SQCD corrections for
the case where a single $b$ quark
is tagged along with the Higgs boson.\footnote{Correction to Higgs plus jet production
in the MSSM from the $gg$ initial state have been examined in 
Refs. \cite{Brein:2003df,Brein:2007da,Brein:2007ej,Field:2003yy}.}
These corrections, along with the NLO QCD corrections previously
computed,\cite{Dittmaier:2003ej,Dawson:2004sh,Campbell:2002zm} and the summation
of the logarithmic threshold corrections\cite{Field:2007ye},
 can then be used to obtain reliable ${\cal O}(\alpha_s^2)$ 
predictions.  

An  effective
Lagrangian approach
has been extensively used in the literature to include SQCD
effects to the $b {\overline b}\phi$ vertex\cite{Carena:1999py,
Hall:1993gn,Carena:2006ai} since these corrections are enhanced for 
large values of $\tan\beta$. The effective Lagrangian is derived assuming
$M_\phi<<M_{SUSY}$, where $M_{SUSY}$ is a
typical squark or gluino mass scale.  We investigate the range of validity
of the effective Lagrangian approach for computing
the SQCD corrections to the $b\phi$ production process.
In addition, we consider the decoupling properties
of our results for heavy MSSM mass scales.  For the decay process, 
$h^0\rightarrow b {\overline b}$, heavy SUSY particles do not decouple 
unless the pseudoscalar mass, $M_A$,
 is also large with respect to the electroweak
scale\cite{Haber:2000kq,Guasch:2003cv} and 
we observe a similar phenomena in our results.  

In Section 2, we discuss the general  MSSM framework of our calculation and 
the
effective Lagrangian approach for approximating the SQCD contributions of squarks
and gluinos.  Section
3 contains the lowest order results for the production process,
$b g\rightarrow b \phi$, a discussion of the renormalization framework
for the SQCD NLO contributions,
and the SQCD NLO results. Numerical results for the Tevatron and the
LHC, along with a discussion of the decoupling properties of
the SQCD contributions are given
in Section 4 and analytic results for the SQCD corrections
 are gathered in an appendix.  Finally, in Section 5,
we present some conclusions.   Section 5 contains predictions for the $b\phi$
cross sections at the Tevatron and the LHC which contain all 
known ${\cal O}(\alpha_s^2)$ 
contributions and thus are the most reliable calculations available.

\section{Framework}
\subsection{MSSM Basics}
The MSSM has been extensively studied in the literature and comprehensive
reviews can be found in Refs. \cite{Djouadi:2005gj,Gunion:1989we}.  
Here we briefly summarize
those aspects of the MSSM relevant for our calculation.

In the MSSM, there are two $SU(2)_L$ Higgs doublets, $H_u$ and $H_d$, which
can be written as\cite{Gunion:1989we,Djouadi:2005gj,Carena:2002es},
\begin{eqnarray}
H_d&=\left(
\begin{array}{c}
h_d^+\\
{1\over \sqrt{2}}(v_1+h_d^0+i\chi_d^0)\\
\end{array}
\right)\, ,\quad
H_u &=\left(
\begin{array}{c}
{1\over \sqrt{2}}(v_2+h_u^0 -i\chi_u^0)\\
-h_u^-
\end{array}
\right)\, .
\end{eqnarray}
After spontaneous symmetry breaking, the $W$ and $Z$ bosons obtain masses and fix
 $v_{SM}^2
=v_1^2+v_2^2=(246~{\rm{GeV}})^2$, while the ratio of the VEVs, $\tan \beta={v_2\over v_1}$, 
is a free parameter of the theory.  There are five physical Higgs
bosons, $h^0,H^0, A^0$ and $H^\pm$, remaining in the theory.  

The scalar potential of the MSSM is described
at tree level by two free parameters, which are usually taken to be $\tan\beta$ and
the mass of the pseudoscalar Higgs boson, $M_A$.  In terms of these parameters, the 
remaining scalar masses, $M_{H^\pm}, M_h$ and $M_H$, are predicted quantities with,
\begin{equation}
  M_{H^\pm}^2=M_A^2+M_W^2\, .
\end{equation}
The physical neutral Higgs bosons, $h^0$ and $H^0$,
 are linear combinations of $h_d^0$
and $h_u^0$,
\begin{equation}
\left(
\begin{array}{c}
h^0\\
H^0
\end{array}
\right)
=
\left(
\begin{array}{cc}
c_\alpha & -s_\alpha\\
s_\alpha & c_\alpha
\end{array}
\right)
\left(
\begin{array}{c}
h_u^0\\
h_d^0
\end{array}
\right)
\label{hmixing}
\end{equation}
where $c_\alpha\equiv\cos\alpha$ and $s_\alpha\equiv\sin\alpha$.  The neutral Higgs boson masses at tree
level are given by,
\begin{equation}
M_{h,H}^2={1\over 2} \biggl\{ M_A^2+M_Z^2\mp
\sqrt{(M_A^2+M_Z^2)^2-4M_Z^2M_A^2\cos^2 2\beta}\biggr\} \, .
\label{treemass}
\end{equation}
Eq. \ref{treemass} implies a tree level upper bound on the lightest Higgs boson mass, $M_h(tree)< M_Z$.
Furthermore, at tree-level,
\begin{equation}
\tan 2 \alpha =\tan 2 \beta\biggl({M_A^2+M_Z^2\over M_A^2-M_Z^2}\biggr)\, .
\label{alpha_def}
\end{equation}
The predictions of Eqs. \ref{treemass} and \ref{alpha_def} receive large 
 radiative corrections of ${\cal O}(G_F m_t^4)$ which raise
the lower bound on $m_h$ to $130-140~{\rm {GeV}}$.\footnote{The upper bound on the
lightest Higgs mass depends on the stop mass, as well as other MSSM parameters.
For a review see Ref. \cite{Djouadi:2005gj}.}  We include these
corrections
using the program FeynHiggs, which generates an effective 
mixing angle, $\alpha_{eff}$, and radiatively corrected 
values for the Higgs boson masses\cite{Hahn:2005cu}.  Using  $\alpha_{eff}$
in the tree level couplings incorporates the bulk of the MSSM 
corrections to the Higgs masses and mixing angles\cite{Heinemeyer:2004gx}.

The MSSM Yukawa couplings are given at tree level by,
\begin{equation}
L_{YUK}=-\lambda_b {\overline \psi_L}
 H_d b_R -\lambda_t {\overline \psi_L} H_u t_R+h.c. \, ,
\label{yukdef}
\end{equation}
where $\psi_L^\dagger=(t_L, b_L)$. Eq. \ref{yukdef} generates masses for the $t$ and
$b$ quarks,
\begin{eqnarray}
m_b&=&{\lambda_b v_1\over \sqrt{2}}
\nonumber \\
m_t&=&{\lambda_t v_2\over \sqrt{2}}\, .
\end{eqnarray}
 We note that the bottom 
quark only couples to $H_d$, while the top quark only couples to $H_u$.
In terms of the physical Higgs mass eigenstates,
\begin{eqnarray}
L_{YUK}&=&-{m_b\over v_{SM}}\biggl( -{s_\alpha\over c_\beta} {\overline b} b h^0
+{c_\alpha \over c_\beta}  {\overline b} b H^0
-t_\beta  {\overline b}i\gamma_5 b A^0\biggr)\nonumber \\
&\equiv & \Sigma g_{b{\overline b}\phi_i}^{LO} {\overline b} b \phi_i\, ,
\label{loyuk}
\end{eqnarray}
where $c_\beta=\cos\beta$, $s_\beta=\sin\beta$, and $t_\beta=\tan\beta$.  We see that the
Yukawa couplings to the $b$ quark are enhanced for large values of $\tan\beta$.

In the MSSM, the scalar partners of the left- and right- handed $b$ quarks, ${\tilde b}_L$
and ${\tilde b}_R$, are not mass eigenstates, but mix according to,
\begin{equation}
L_M=-({\tilde b}^*_L, {\tilde b}^*_R)M_{\tilde b}^2 \left(
\begin{array}{c}
{\tilde b}_L \\
{\tilde b}_R
\end{array}
\right)\, .
\end{equation}
 The ${\tilde b}$ squark  mass matrix is,
\begin{equation}
M_{{\tilde b}}^2=\left(
\begin{array}{cc}
{\tilde m}_L^2 & m_b (A_b-\mu\tan\beta)\\
m_b(A_b-\mu\tan\beta) & {\tilde m}_R^2\\
\end{array}
\right)\, ,
\label{squarkm}
\end {equation}
where we define,
\begin{eqnarray}
{\tilde m}^2_L&=&
{ M}_Q^2 +m_b^2+M_Z^2\cos 2\beta (I_3^b-Q_b\sin^2\theta_W)\nonumber \\
{\tilde m}^2_R&=&
{M}_D^2 +m_b^2+M_Z^2\cos 2\beta Q_b\sin^2\theta_W\, ,
\end{eqnarray}
and ${ M}_{Q,D}$ are the soft SUSY breaking masses,
$I_3^b=-1/2$, and $Q_b=-1/3$.  The parameter $A_b$ is the trilinear scalar coupling of the soft supersymmetry breaking Lagrangian and $\mu$ is the Higgsino mass parameter.
The $b$ squark mass eigenstates are
${\tilde b}_1$ and ${\tilde b}_2$ and define
 the $b$-squark mixing angle, ${\tilde\theta_b}$
\begin{eqnarray}
{\tilde b}_1&=& \cos {\tilde\theta_b} {\tilde b}_L 
+\sin{\tilde\theta_b} {\tilde b}_R 
\nonumber \\
{\tilde b}_2&=& -\sin{\tilde\theta_b}
 {\tilde b}_L +\cos{\tilde\theta_b} {\tilde b}_R\, , 
\nonumber \\
\end{eqnarray}
where at tree level, 
\begin{equation}
\tan 2 {\tilde\theta_b}={m_b(A_b-\mu \tan\beta)\over {\tilde m}_L^2-
{\tilde m}_R^2}\, .
\end{equation}
Eq. \ref{squarkm} gives the squark mass eigenstates at tree level,
\begin{equation}
 m_{{\tilde b}_{1,2}}^2=\biggl( {\tilde m}_L^2\cos^2\tilde{\theta}_b
+{\tilde m}_R^2 \sin^2\tilde{\theta}_b\biggr) \mp 
\sin (2{\tilde \theta}_b)m_b(A_b-\mu\tan\beta)\, .
\end{equation}

The  $b$ squarks (unlike the $b$ quark) couple to both Higgs
doublets, $H_u$ and $H_d$ and the Feynman rules for the squark-squark-Higgs
couplings can be found in the appendices of Ref. \cite{Gunion:1989we}.
Finally, we use the Feynman diagramatic techniques of Ref \cite{Denner:1992vza}
 and the
Feynman rules for the squark/gluino interactions which are
 given in Appendix A of Ref. \cite{Berge:2007dz}.

\subsection{Effective Lagrangian Approach}

\begin{figure}[t]
\begin{center}
\includegraphics[scale=0.8]{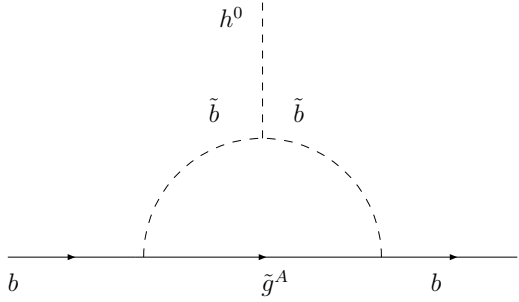} 
\caption[]{SUSY QCD contribution from gluinos and squarks 
to the $b{\overline b}h^0$ 
effective vertex. Both squarks, ${\tilde b}_1$ and  ${\tilde b}_2$, contribute
to the interaction. }
\label{fg:virta}
\end{center}
\end{figure}
At tree level, there is no $ {\overline \psi}_L b_R H_u$
 coupling in the MSSM (see Eq. \ref{yukdef}), but such a 
coupling arises at one loop through the
Feynman diagram shown in Fig. \ref{fg:virta} and gives rise to an effective
interaction\cite{Carena:1999py,Hall:1993gn},
\begin{equation}
L_{eff}=-\lambda_b
{\overline \psi}_L\biggl(H_d+{\Delta m_b\over \tan\beta}
H_u\biggr)b_R+h.c. \,\,\quad .
\label{effdef}
\end{equation}
Eq. \ref{effdef} shifts the $b$ quark mass from its tree level value,
\begin{equation}
m_b={\lambda_b v_1\over \sqrt{2}} (1+\Delta m_b)\, ,
\end{equation}
and  also implies that the Yukawa couplings of the
Higgs bosons to the $b$ quark are shifted from the
tree level predictions.  The shift of the
Yukawa couplings  can be included
with an effective Lagrangian approach\cite{Carena:1999py,Guasch:2003cv},
\begin{eqnarray}
L_{eff}&=&-{m_b\over v_{SM}}\biggl({1\over 1+\Delta m_b}\biggr)
\biggl(-{\sin \alpha \over \cos\beta}\biggr)\biggl(1-{\Delta m_b\over \tan\beta
\tan \alpha}\biggr) {\overline b} b h^0\nonumber \\
&-&{m_b\over v_{SM}}\biggl({1\over 1+\Delta m_b}\biggr)
\biggl({\cos \alpha \over \cos\beta}\biggr)\biggl(1+{\Delta m_b \tan\alpha
\over \tan\beta}\biggr) {\overline b} b H^0\nonumber \\
&-&{m_b\over v_{SM}}\biggl({1\over 1+\Delta m_b}\biggr)
\biggl(-\tan\beta\biggr)\biggl(1-{\Delta m_b\over \tan^2\beta}\biggr) 
{\overline b}i\gamma_5 b A^0\nonumber \\
&\equiv & g_{b{\overline b}h} {\overline b} b h^0
+g_{b{\overline b}H} {\overline b} b H^0
+g_{b{\overline b}A} {\overline b}i\gamma_5 b A^0 \, .
\label{mbdef}
\end{eqnarray}
The Lagrangian of Eq. \ref{mbdef} has been shown to
 sum all terms of ${\cal O}(\alpha_s^n\tan^n\beta)$ for large 
$\tan\beta$\cite{Carena:1999py}.\footnote{It is also possible to sum the 
contributions which are proportional to $A_b$, but these terms are
less important numerically\cite{Guasch:2003cv}.}
This effective Lagrangian has
been used to compute both the inclusive 
production process, $b {\overline b}
\rightarrow h^0$, and the decay process, $h^0\rightarrow b {\overline b}$,
and  yields results which are within a few percent of the exact
one-loop SQCD calculations\cite{Guasch:2003cv,Dittmaier:2006cz}.

The expression for $\Delta m_b$ is found in the limit
 $m_b << m_\phi, M_Z <<m_{{\tilde b}_1}, m_{{\tilde b}_2}, m_{\tilde g}$ .  The
contribution to $\Delta m_b$ from sbottom/gluino loops 
is\cite{Carena:1994bv,Hall:1993gn,Carena:1999py} 
\begin{equation}
\Delta m_b={2\alpha_s(\mu_R)\over 3 \pi} m_{\tilde g} \mu 
\tan\beta
I(m_{\tilde {b_1}},
m_{\tilde{ b_2}}, m_{\tilde g})\, ,
\label{db}
\end{equation}
where the function $I(a,b,c)$ is,
\begin{equation}
I(a,b,c)={1\over (a^2-b^2)(b^2-c^2)(a^2-c^2)}\biggl\{a^2b^2\log\biggl({a^2\over b^2}\biggr)
+b^2c^2\log\biggl({b^2\over c^2}\biggr)
+c^2a^2\log\biggl({c^2\over a^2}\biggr)\biggr\}\, ,
\end{equation}
and $\alpha_s(\mu_R)$ should be evaluated at a typical squark or gluino mass. 
Note that Eq. \ref{db} is valid for arbitrary values of $\tan\beta$.
There is also an electroweak correction to $\Delta m_b$ which we neglect,
since we are concerned only with the ${\cal O}(\alpha_s^2)$ contributions
in this paper.

Eq. \ref{db} is a non-decoupling effect in the sense that if  the 
 mass scales of the squarks and gluino, along with the mixing
parameter $\mu$, become
large for fixed $M_A$, $\Delta m_b$ does not vanish, 
\begin{equation}
\Delta m_b\rightarrow -sign(\mu){\alpha_s\over 3 \pi} \biggl(\tan\beta
+\cot \alpha\biggr)\, .
\label{decouple}
\end{equation}
In the large $M_A$ limit, 
\begin{equation}
\tan\beta +\cot\alpha \rightarrow -{2 M_Z^2\over M_A^2}\tan\beta \cos 2\beta
+{\cal O}\biggl({M_Z^4\over M_A^4}\biggr)\, ,
\label{decma}
\end{equation}
and the decoupling limit of the MSSM is recovered\cite{Haber:2000kq}. 

In the next sections, we investigate the use of the effective Lagrangian
of Eq. \ref{mbdef} to estimate the SQCD corrections to the process $gb\rightarrow
b \phi$ and compare the results with the complete ${\cal O}(\alpha_s^2)$ 
one-loop SQCD
calculation.

\section{$gb\rightarrow b\phi$ at NLO in SUSY QCD}
In this section, we summarize the expressions for the SQCD corrections to the
$g b\rightarrow b \phi$ process.

\subsection{Lowest Order}
\begin{figure}[t]
\begin{center}
\includegraphics[scale=0.8]{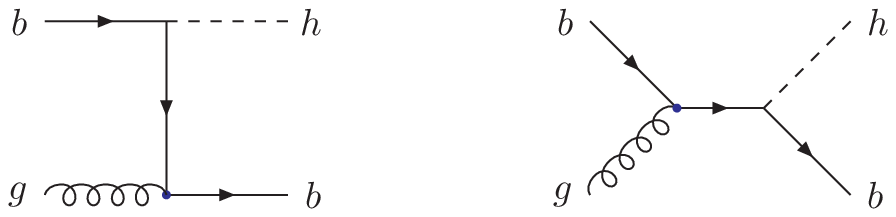} 
\caption[]{Feynman diagrams for $ g(q_1)+b (q_2)\rightarrow
b(p_b)+ \phi_i(p_h)$, where $\phi_i=h^0, H^0$ or $A^0$.}
\label{fg:bghb_feyn}
\end{center}
\end{figure}

  The tree level diagrams for $g(q_1) + b(q_2) \to b(p_b) + \phi(p_h)$
 are shown 
in Fig. \ref{fg:bghb_feyn}.
The lowest order amplitude can be written
as a linear sum over two allowed kinematic structures,
with a third contributing at NLO,\footnote{For pseudoscalar
production, the three kinematic structures of Eq. \ref{kin} include
a $\gamma_5$.}
\begin{eqnarray}
\label{eq:Ms}
M_s^\mu &=& \bar{u}(p_b) ({\not{q_1}} + {\not{q}}_2) \gamma^\mu u(q_2) 
\nonumber \\
\label{eq:Mt}
M_t^\mu &=& \bar{u}(p_b) \gamma^\mu ({\not{p}}_b - {\not{q}}_1) u(q_2)
\nonumber \\
\label{eq:M1}
M_1^\mu &=& q_2^\mu\bar{u}(p_b) u(q_2)  \, .
\label{kin}
\end{eqnarray}
The Born level amplitude is
\begin{equation}
A_{Born}^\mu=g_s T^a g_{b\bar{b} \phi_i}^{LO}    
A_{LO}^\mu\, ,
\end{equation}
where,
\begin{equation}
\label{eq:ampLOs}
A_{LO}^\mu = 
  \biggl(\frac{M_s^\mu}{s}+\frac{M_t^\mu}{t}\biggr)\,,
\end{equation}
and we define the usual Mandelstam variables as:
\begin{eqnarray}
s &\equiv& (q_1 + q_2)^2 = (p_b + p_h)^2 \,,\nonumber \\
t &\equiv& (q_1 - p_b)^2 = (q_2 - p_h)^2 \,,\nonumber \\
u &\equiv& (q_1 - p_h)^2 = (q_2 - p_b)^2 \,.
\end{eqnarray}
The spin and color averaged tree level partonic cross section is,
\begin{equation}
{d{\hat\sigma}_{Born}\over dt}(bg\rightarrow b\phi)
={\alpha_s(\mu_R)\over 96  s^2}\biggl( g_{b{\overline b} \phi}^{LO}\biggr)^2
 \mid A_{LO}^\mu\mid^2\, .
\label{siglo}
\end{equation}
The lowest order, ${\cal O}(\alpha_s)$, hadronic cross section is found in the usual manner by
integration with the $b$ quark and gluon PDFs,
\begin{equation}
\sigma_{LO}(pp\rightarrow b \phi)=\int dx_1 dx_2 \biggl[
g(x_1,\mu_F)b(x_2,\mu_F)
\int dt \biggl({d{\hat \sigma}_{Born}\over dt}\biggr)
+(x_1\leftrightarrow x_2)\biggr]\, .
\label{lowsig}
\end{equation}

\subsection{NLO QCD Corrections}
The NLO pure QCD ${\cal O}(\alpha_s^2)$ corrections to the $g b\rightarrow b \phi$ process consist of one-loop virtual  corrections containing
$b$ quarks and gluons and
real gluon emission diagrams, along with appropriate counterterms.
These corrections have been computed by several groups with excellent agreement
between the groups\cite{Dawson:2005vi,Dawson:2003kb,Campbell:2004pu,Maltoni:2003pn,Dawson:2004sh,Dittmaier:2003ej,Dicus:1998hs,Campbell:2002zm}.
  Schematically, 
\begin{eqnarray}
\sigma_{NLO}(pp\rightarrow b\phi)_{QCD}& \equiv & \biggl({g_{b{\overline b}\phi}\over
g^{LO}_{b{\overline b}\phi}}\biggr)^2
\sigma_{LO}(pp\rightarrow b\phi)
+\delta\sigma_{QCD} (pp\rightarrow b\phi)
\nonumber \\
&\equiv & \sigma_{NLO}({\hbox{gluon~only}})\, .
\label{qcdsig}
\end{eqnarray}
In Eq. \ref{qcdsig}, we normalize the Yukawa couplings
for  both the Born contribution
to the NLO cross section  and for $\delta\sigma_{QCD}$ with
$g_{b {\overline b}\phi}$  (defined by the effective 
Lagrangian of Eq. \ref{mbdef}).  This algorithm for including the
Yukawa couplings of Eq. \ref{mbdef} in the  NLO QCD corrections corresponds to 
the convention of Ref. \cite{Dawson:2005vi} and so the  NLO QCD corrections presented
in Section 4 (and labelled ``NLO (gluon only)'') can be directly compared
with this reference.

\subsection{Improved Born Approximation}

One of the major
goals of this work is to investigate the accuracy of the Improved Born 
Approximation (IBA). In order to make the comparison as transparent as possible,
we define an Improved Born Approximation in which the Born amplitude is
normalized by the Yukawa couplings, $g_{b {\overline b}\phi}$, of 
Eq. \ref{mbdef},
\begin{equation} {d{\hat{\sigma}}_{IBA}\over dt}
\equiv {d{\hat{\sigma}}_{Born}\over dt}\biggl({g_{b {\overline b}\phi}
\over
g_{b {\overline b}\phi}^{LO}}\biggr)^2\, .
\label{sigiba}
\end{equation}
The Improved Born Approximation  incorporates the 
effective Lagrangian approximation to the  SQCD effects
on the $b {\overline b} \phi$ Yukawa couplings at low
energy, but does not include the full SQCD calculation presented in the
next section.  In particular, the ``Improved Born Approximation'' does not
include contributions from box diagrams including internal squarks and gluinos or 
the full momentum dependence of the SQCD contributions.

\subsection{One Loop SQCD Corrections}

The amplitudes for the SQCD one-loop corrections from sbottom/gluino
loops can be written as:
\begin{equation}
\label{eq:ampNLO}
{\cal{A}}_i^\mu = \frac{ \alpha_s}{4 \pi} (g_s T^a) \biggl[
  X_i^{(s)} M_s^\mu + X_i^{(t)} M_t^\mu + X_i^{(1)} M_1^\mu \biggr]\,,
\end{equation}
where the matrix elements $M_s, M_t$ and $M_1$ are given in Eq. \ref{kin}
and expressions for the individual contributions, $X_i$  are found in
 Appendix B. 
The integrals of Appendix A are evaluated
in $n=4-2\epsilon$ dimensions.
The diagrams consist of self-energy, vertex, and box contributions,
along with the appropriate counterterms to cancel the ultraviolet
divergences. We consistently neglect contributions which are
suppressed by terms of ${\cal O}(m_b^2/s,m_b^2/M_\phi^2)$, etc.
The spin and color averaged partonic cross section to NLO in SQCD is,
\begin{eqnarray}
{d{\hat\sigma}\over dt}(bg\rightarrow b\phi)_{SQCD}
&_= &
{d{\hat\sigma}\over dt}(bg\rightarrow b\phi)\mid_{IBA}
+{\alpha_s^2(\mu_R)\over 192 \pi s^2} 
 \biggl(g_{b{\overline b}\phi}\biggr)^2
\sum_i|A_{LO} \cdot \biggl(X_i^{(s)}M_s^*+
\nonumber \\
&& X_i^{(t)}M_t^*+X_i^{(1)}M_1^*\biggr)+{d{\hat\sigma}_{CT}\over dt}\, ,
\label{signlo}
\end{eqnarray}
where the contribution from the counterterms is  discussed in the
next section.  
The SQCD NLO contribution to the hadronic cross section is
\begin{equation}
\sigma_{NLO}(pp\rightarrow b \phi)_{SQCD}=\int dx_1 dx_2 \biggl[
g(x_1,\mu)b(x_2,\mu)
\int dt \biggl({d{\hat \sigma}_{SQCD}\over dt}\biggr)
+(x_1\leftrightarrow x_2)\biggr]\, .
\label{sqcdsig}
\end{equation}
The result of Eq. \ref{sqcdsig} is labelled ``NLO (gluino/squark only)''
in the figures of Section IV.

Finally, we note that
\begin{eqnarray}
|A_{LO} \cdot M_s^*| &=& -4 
  \biggl[ \frac{t m_\phi^2 + u^2 + u s}{t} \biggr] 
\nonumber\\
|A_{LO} \cdot M_t^*| &=& -4 
  \biggl[ \frac{s m_\phi^2 - u s + u m_\phi^2}{s} \biggr] 
\nonumber \\
|A_{LO} \cdot M_1^*| &=& -2 
  \biggl[ \frac{u^2 + u s}{ t} \biggr] \, .
\end{eqnarray}

\subsection{Counterterms}

In this section, we discuss the squark and gluino contributions to the counterterms
which are necessary for the SQCD one-loop calculation.  The renormalization of the pure QCD
contribution has been presented previously\cite{Dawson:2005vi,Dawson:2003kb,Campbell:2004pu,Maltoni:2003pn,Dawson:2004sh,Dittmaier:2003ej,Dicus:1998hs,Campbell:2002zm}.

Self-energy and vertex corrections to the tree-level process $gb \to b\phi$ process give
rise to ultraviolet (UV) divergences.  These singularities are cancelled by a set of 
counterterms fixed by well-defined renormalization conditions.  The renormalization of the
propagators and interaction vertices of the theory reduces to introducing
counterterms for the external field wavefunctions of the bottom quarks ($\delta Z_V$) and
gluons ($\delta Z_3$),  and for the strong 
coupling constant ($\delta Z_g$).  The counterterm for the bottom quark Yukawa coupling,
$g_{b\bar{b}\phi}^{LO} \sim m_b/v$, coincides with the counterterm for the bottom quark mass,
since the vacuum expectation value $v$ is not renormalized at the one-loop level of 
SUSY QCD.  (We follow the approach of Refs. \cite{Berge:2007dz,Hafliger:2005aj} for the SQCD renormalization).

\begin{figure}[t]
\begin{center}
\includegraphics[scale=0.8]{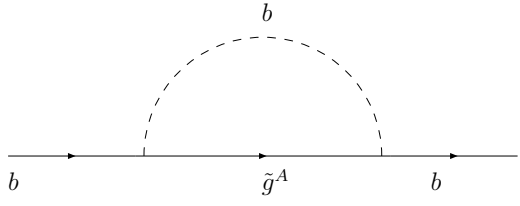} 
\caption[]{SUSY QCD contribution to the $b$ quark self energy. Both squarks, ${\tilde b}_1$
and ${\tilde b}_2$, contribute to the interaction.}
\label{fg:selfen}
\end{center}
\end{figure}

We define the $b$ quark self energy as
\begin{eqnarray}
\Sigma^b(p) & =&{\not p}\biggl(\Sigma_V^b(p^2)-\Sigma_A^b(p^2) \gamma_5\biggr)
+m_b\Sigma_S^b(p^2)\nonumber \\
\delta\Sigma^b(p) &=& {\not p}\biggl(\delta Z_V^b-\delta
 Z_A^b\gamma_5\biggr)-m_b\delta Z_V^b
-\delta m_b\, ,
\end{eqnarray}
which yields the renormalized propagator,
$\Sigma^b_{ren}(p)=\Sigma^b(p)+\delta\Sigma^b(p)$,
\begin{equation}
\Sigma^b_{ren}(p)=({\not p}-m_b)\biggl(\Sigma_V^b+\delta Z_V^b\biggr)
+m_b\biggl(\Sigma_V^b+\Sigma_S^b-{\delta m_b\over m_b}\biggr) \, .
\end{equation}
We evaluate the SQCD contribution to the $b$ mass using 
the on-shell renormalization condition,
\begin{eqnarray}
\Sigma^b({\not p}=m_b)&=& 0
\nonumber \\
lim_{{\not {p}}\rightarrow m_b} {\Sigma^b(p)\over {\not {p}}-m_b}&=& 0\, .
\end{eqnarray}
This scheme decouples the gluino and $b$-squark from the running of the $b$ 
quark Yukawa coupling. 
The self energy contribution from gluino-squark loops is shown in 
Fig. \ref{fg:selfen}
and yields the on-shell result,
\begin{eqnarray}
{\delta m_b\over m_b}
&=&\biggl(\Sigma_V^b+\Sigma_S^b\biggr)\mid_{p^2=m_b^2}
\nonumber \\
&=&-{\alpha_s(\mu_R)\over 3 \pi}\biggl\{ B_1(m_b^2, m_{\tilde g}, 
 m_{\tilde b_1})
 +B_1(m_b^2, m_{\tilde g}, m_{\tilde b_2})\nonumber \\
&&\qquad +2\sin {\tilde \theta_b}
{\tilde \cos\theta_b} {m_{\tilde g}\over m_b}
\biggl[  B_0(m_b^2, m_{\tilde g}, m_{\tilde b_1})-
 B_0(m_b^2, m_{\tilde g},  m_{\tilde b_2})\biggr]\biggr\}\, ,
\end{eqnarray}
where the integrals $B_0$ and $B_1$ are defined in Appendix A.

The $b$-quark self-energy is renormalized by $\Psi\rightarrow  
\sqrt{Z_V^b}\Psi=\sqrt{1+\delta Z_V^b}\Psi$.
Neglecting contributions suppressed by powers of $m_b$, in the on-shell
scheme we find,
\begin{eqnarray}
\delta Z_V^b
&=&-\Sigma_V\mid_{p^2=m_b^2}
\nonumber \\
&=&{\alpha_s(\mu_R)\over 3 \pi}\biggl\{ B_1(m_b^2, m_{\tilde g}, 
m_{\tilde b_1})
 +B_1(m_b^2, m_{\tilde g}, m_{\tilde b_2})\biggr\}\, .\nonumber \\
\end{eqnarray}

The external gluon has the wavefunction renormalization, $g_\mu^A
\rightarrow \sqrt{Z_3} g_\mu^A=\sqrt{1+\delta Z_3}g_\mu^A$ and 
the renormalization of the strong coupling is found from $g_s
\rightarrow Z_g g_s$.  We have further, $\delta Z_g=-\delta Z_3/2$.\footnote{The validity of
this relation for the squark/gluino contributions to the SQCD renormalization is 
demonstrated in Ref. \cite{Berge:2007dz} and references therein.}
We renormalize $g_s$ using the ${\overline{MS}}$ scheme modified to
decouple heavy SUSY particles, ${\it i.e.}$ the heavy squark and gluino contributions
are evaluated at zero momentum\cite{Nason:1987xz},
\begin{eqnarray}
\delta Z_3 &=& -{\partial \Sigma_T(p^2)\over \partial p^2}\mid_{pole}
\nonumber \\
&=& -{\alpha_s(\mu_R)\over 4 \pi}\biggl\{ 
{1\over 6} 
\Sigma_{{\tilde q_i}}
\biggl(
{4 \pi \mu_R^2\over m_{{\tilde q}_i}^2}
\biggr)^\epsilon +
2\biggl(
{4 \pi \mu_R^2\over m_{\tilde g}^2}\biggr)^\epsilon \biggr\} 
{1\over \epsilon}\Gamma(1+\epsilon)\,,
\label{z3def}
\end{eqnarray}
where $\Sigma_T$ represents the transverse portion of the gluon self energy, 
the sum in Eq. \ref{z3def} is over all squarks (not just the $b$ squark),
and $\mu_R$ is an appropriate renormalization scale.

The Yukawa couplings are defined by Eq.\ref{mbdef}, which includes the summation
of large $\tan\beta$ effects. Using Eq. \ref{mbdef} for the Yukawa 
couplings of the Born contribution in Eq. \ref{signlo}
includes some one-loop effects in the first term
 which must be subtracted   in order not to double count.
This generates 
additional counterterms,
\begin{eqnarray}
{{\delta \tilde  m_b}^h\over m_b}& =&\Delta m_b\biggl(1+{1\over \tan\alpha \tan\beta}
\biggr) \qquad (h{\hbox{~production)}}\nonumber \\
{{\delta \tilde m_b}^H\over m_b}& =&\Delta m_b\biggl(1-{\tan\alpha \over \tan\beta}
\biggr) \qquad (H{\hbox{~production)}}\nonumber \\
{{ \delta\tilde  m_b}^A\over m_b}& =&\Delta m_b\biggl(1+{1\over \tan^2\beta}
\biggr) \qquad (A{\hbox{~production)}}\, .
\end{eqnarray}

The counterterms make a contribution to the total partonic SQCD cross section,
\begin{equation}
{d {\hat \sigma_{CT}}\over dt}
= {d {\hat \sigma}_{IBA}\over dt}
\biggl(2\delta Z_V^b+ \delta Z_3 + 2\delta Z_g
+2 {\delta m_b\over m_b}+2 {\delta \tilde m_b^\phi
\over m_b}\biggr) \, .
\label{ctdef}
\end{equation}
 We 
note that the $1/\epsilon$ contributions cancel in Eq. \ref{ctdef}, although
there are remaining finite pieces.

\subsection{Complete NLO Result}
The complete ${\cal O}(\alpha_s^2)$ 
NLO prediction includes both the pure QCD NLO corrections
and the squark/gluino NLO SQCD contributions, along with the
Born contribution\footnote{Note the the Born contribution is (arbitrarily for
this purpose) assigned to $\sigma_{NLO}(pp\rightarrow b\phi)_{SQCD}$} ,
\begin{equation}
\sigma_{NLO}(pp\rightarrow b\phi)_{QCD+SQD}\equiv
\sigma_{NLO}(pp\rightarrow b\phi)_{SQCD}+
+\delta\sigma_{QCD}\, . 
\label{comdef}
\end{equation}  
The curves labelled ``Complete NLO'' in Section IV correspond to Eq. \ref{comdef}
and represent the most accurate predictions available for the $pp\rightarrow
b\phi$ MSSM cross sections.

\section{Results and Discussion}
Our NLO numerical results are obtained with CTEQ6M parton 
distributions\cite{Lai:1999wy},
 $\alpha_s(\mu_R)$ evaluated with the $2-$loop evolution and 
$\alpha_s^{NLO}(M_Z)=0.118$, and use the Yukawa couplings
of Eq. \ref{mbdef}.  The lowest order  cross sections use 
CTEQ6L PDFs, $\alpha_s(\mu_R)$ evaluated at $1$-loop, and have the lowest order
Yukawa couplings of Eq. \ref{loyuk}.
We require that the outgoing $b$ quark has $p_T>20$ GeV
and pseudorapidity $\mid \eta \mid < 2.0$ for the Tevatron and 
$\mid \eta \mid < 2.5$
for the LHC. In the NLO QCD real gluon emission contribution
(obtained from Ref. \cite{Dawson:2004sh}), the final state gluons
and $b$ quarks are merged into a single jet if their
pseudorapidity/azimuthal separation, $\Delta R=
\sqrt{(\Delta\eta)^2+(\Delta \phi)^2}$, is less than $0.4$. The renormalization/factorization
scales, $\mu_R,\mu_F$, are  taken to be $M_\phi/4$. Finally, the results
labelled ``Improved Born Approximation'' (Eq. \ref{sigiba}), use NLO PDFS
and $\alpha_s(\mu_R)$ evaluated with the $2-$loop evolution and 
$\alpha_s^{NLO}(M_Z)=0.118$.

The $b$ quark mass appearing in the  Yukawa couplings
of Eqs.  \ref{loyuk} and \ref{mbdef} is taken to be the running $\overline {MS}$
$b$ quark mass and is  evaluated at two loops for the NLO predictions 
and the IBA predictions  and at $1-$loop for the LO predictions,
\begin{eqnarray}
\label{eq:mb_ms_1l_2l}
\overline{m}_b(\mu_R)_{1l} &=& m^{\text{pole}}_b
\biggl[ \frac{\alpha_s(\mu_R)}{\alpha_s(m^{\text{pole}}_b)}
\biggr]^{c_0 / b_0}\,\,\,, \nonumber \\
\overline{m}_b(\mu_R)_{2l} &=& m^{\text{pole}}_b
\left[ \frac{\alpha_s(\mu_R)}{\alpha_s(m^{\text{pole}}_b)}
\right]^{c_0 / b_0}
\left[ 1+\frac{c_0}{b_0} \biggl( c_1 - b_1 \biggr)
         \frac{\alpha_s(\mu_R)-\alpha_s(m^{\text{pole}}_b)}{\pi}\right]\,\,\,,
\end{eqnarray}
where, 
\begin{equation}
\label{eq:b0_b1_c0_c1}
\begin{array}{ll}
  b_0 =\frac{1}{4\pi}
        \biggl( \frac{11}{3}N_c - \frac{2}{3} N_{fl} \biggr),\qquad &
  b_1  =\frac{1}{2\pi}
        \biggl( \frac{51 N_c - 19 N_{fl}}{11 N_c - 2 N_{fl}} \biggr), \qquad
\nonumber \\ 
c_0  =\frac{1}{\pi}, &
  c_1  =\frac{1}{72\pi}
        \biggl( 101 N_c - 10 N_{fl} \biggr)\,\,\,,
\end{array}
\end{equation} with $N_c=3$ and
$N_{fl}=5$, the number of light flavors.  We take the
 $b$ quark pole mass, $m_b^{\textrm{pole}}=4.62$~GeV.

The MSSM parameters are found using FeynHiggs to generate an effective Higgs
mixing angle, $\alpha_{eff}$, and radiatively corrected Higgs masses.  
 The MSSM parameters are listed in the figure
captions. The gluino mass is always identified with $M_{SUSY}$, while
the  $b$ squark masses and the  $b$ squark mixing angle,  ${\tilde \theta}_b$,
are given for representative parameter values in Tables \ref{tab:masses} and
\ref{tab:masses2}.

\begin{table}[pt]
\caption{$b$ squark masses and mixing angles from
\cite{Hahn:2005cu}. All soft-SUSY breaking masses are 
taken equal to  $M_{SUSY}= m_{\tilde g}=A_b=A_t$ and
$\mu=M_2=200$~GeV.}
{\begin{tabular}{@{}cccc@{}}\toprule
\multicolumn{4}{c}{$\tan\beta=10$} \\ \hline 
$M_{SUSY}$(TeV)  & ${\tilde \theta}_b$ & $m_{{\tilde b}_1}$ (TeV) &
$m_{{\tilde b}_2}$ (TeV)   \\ \hline 
1 & -0.64 & 1.0 & 1.0 \\
2 &  0.00 & 2.0 & 2.0\\
3 &  0.64 & 3.0 & 3.0\\
4 &  0.71 & 4.0 &4.0\\
5 &  0.74 & 5.0 & 5.0\\
\botrule
\end{tabular}}
\label{tab:masses}
\end{table}

\begin{table}[pt]
\caption{$b$ squark masses and mixing angles from
\cite{Hahn:2005cu}. All soft-SUSY breaking masses are 
taken equal to  $M_{SUSY}= m_{\tilde g} =A_b=A_t$ and
$\mu=M_2=200$~GeV.}
{\begin{tabular}{@{}cccc@{}}\toprule
\multicolumn{4}{c}{$\tan\beta=40$} \\ \hline 
$M_{SUSY}$(TeV)  & ${\tilde \theta}_b$ & $m_{{\tilde b}_1}$ (TeV) &
$m_{{\tilde b}_2}$ (TeV)   \\ \hline 
1 & -0.76 & 1.0 & .98 \\
2 &  -0.76 & 2.0 & 2.0\\
3 &  -0.75 & 3.0 & 3.0\\
4 &  -0.75 & 4.0 &4.0\\
5 &  -0.73 & 5.0 & 5.0\\
\botrule
\end{tabular}}
\label{tab:masses2}
\end{table}

\begin{figure}[hbtp!]
\begin{center}
\begin{tabular}{rl}
\hskip -.2in
\includegraphics[bb=8 24 700 700,scale=0.35]
{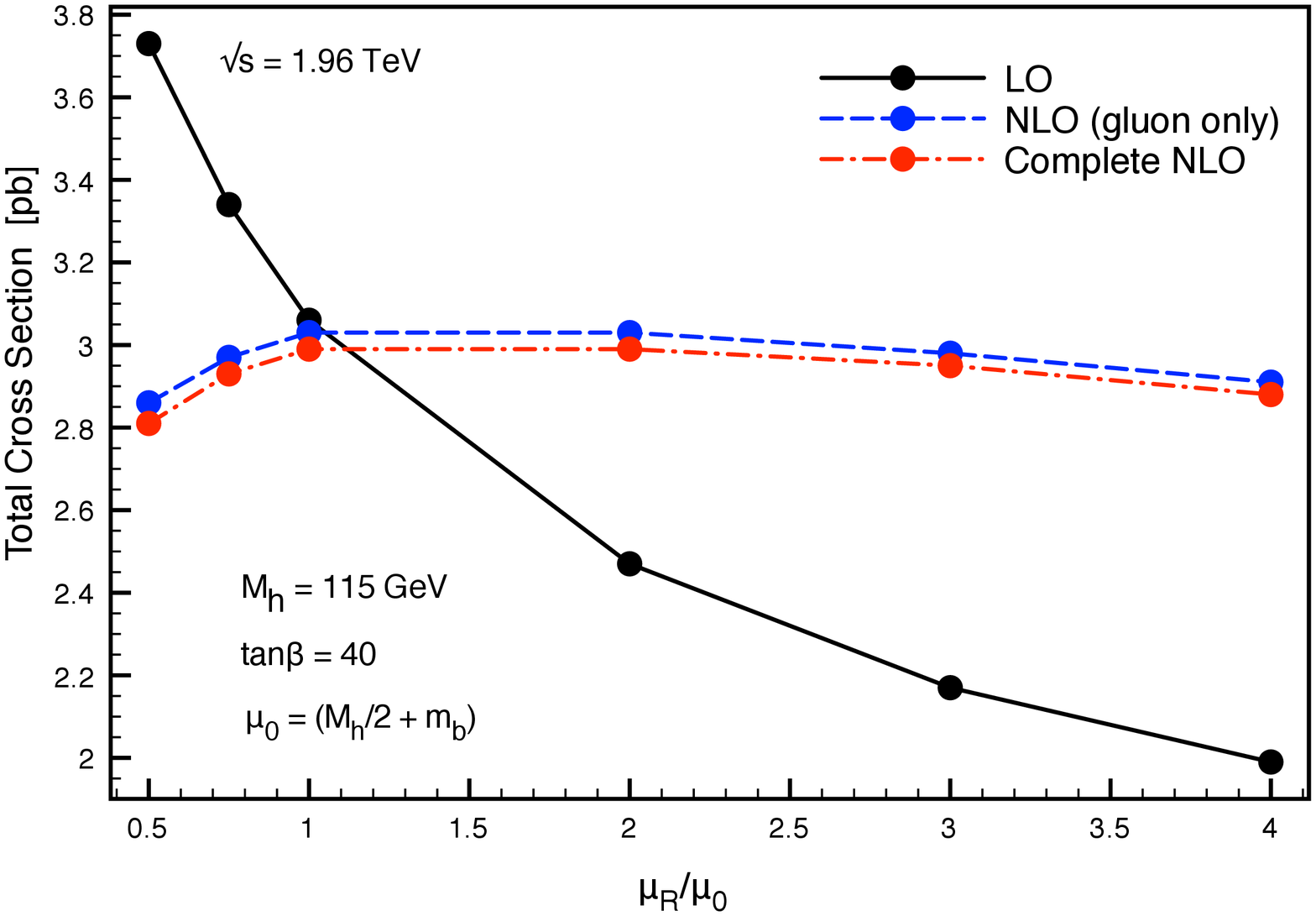} & \hskip .25in
\includegraphics[bb=8 24  700 700,scale=0.35]
{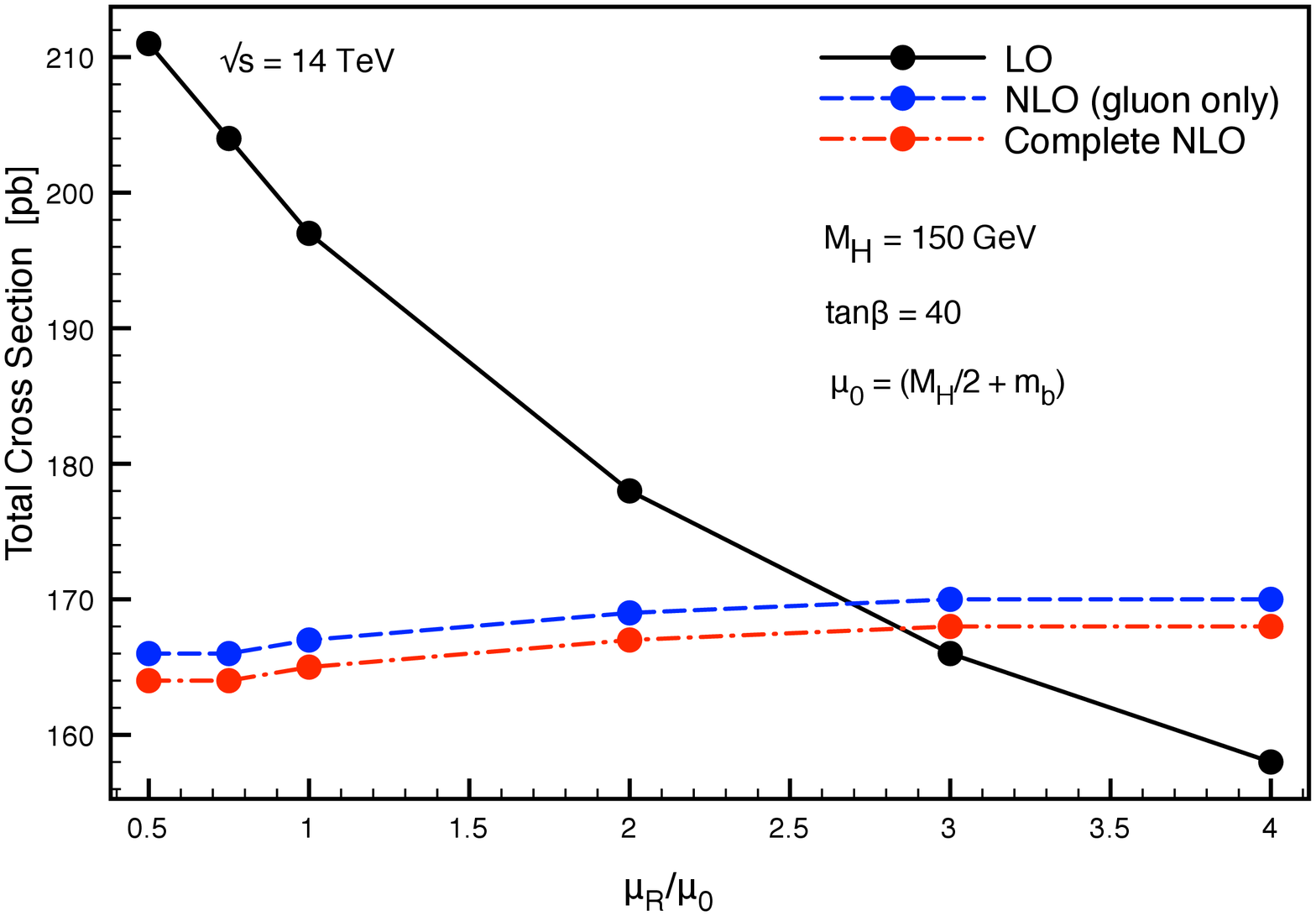}
\end{tabular}
\vspace*{8pt}
\caption[]{Scale dependence of the NLO results for 
$p \overline {p}\rightarrow bh^0$ (LHS)
and $pp\rightarrow bH^0$ (RHS) for $\tan\beta=40$. We set the renormalization/factorization scales equal, $\mu_R=\mu_F$.}  
\label{fg:mudep} 
\end{center}
\end{figure}
The NLO results for the $pp (p{\overline p})
\rightarrow b\phi$ processes depend on both
the renormalization and factorization scales.  For simplicity, we take these
scales equal, $\mu_R=\mu_F$.  Fig. \ref{fg:mudep} shows the 
dependence on scale ($\mu_R/\mu_0$) of the NLO results at both the
Tevatron and the LHC and
demonstrates the expected improvements between the lowest order results and
the NLO results.  The curves labelled ``Complete NLO'' include  the full set
of
QCD and SQCD contributions (Eq. \ref{comdef}) 
and are significantly less dependent on
the choice of scale than the lowest order result (Eq. \ref{lowsig}). 
The dashed curve
(``NLO (gluon only)'', Eq. \ref{qcdsig}) 
only includes the  SQCD contributions via the effective
Yukawa couplings of Eq. \ref{mbdef} and has an almost
identical scale dependence as the full result of Eq. \ref{comdef}.
For the remainder
of our plots, we choose the renormalization/factorization
scale to be $M_\phi/4$.

\begin{figure}[hbtp!]
\begin{center}
\begin{tabular}{rl}
\hskip -.2in
\includegraphics[bb=8 24 700 700,scale=0.35]
{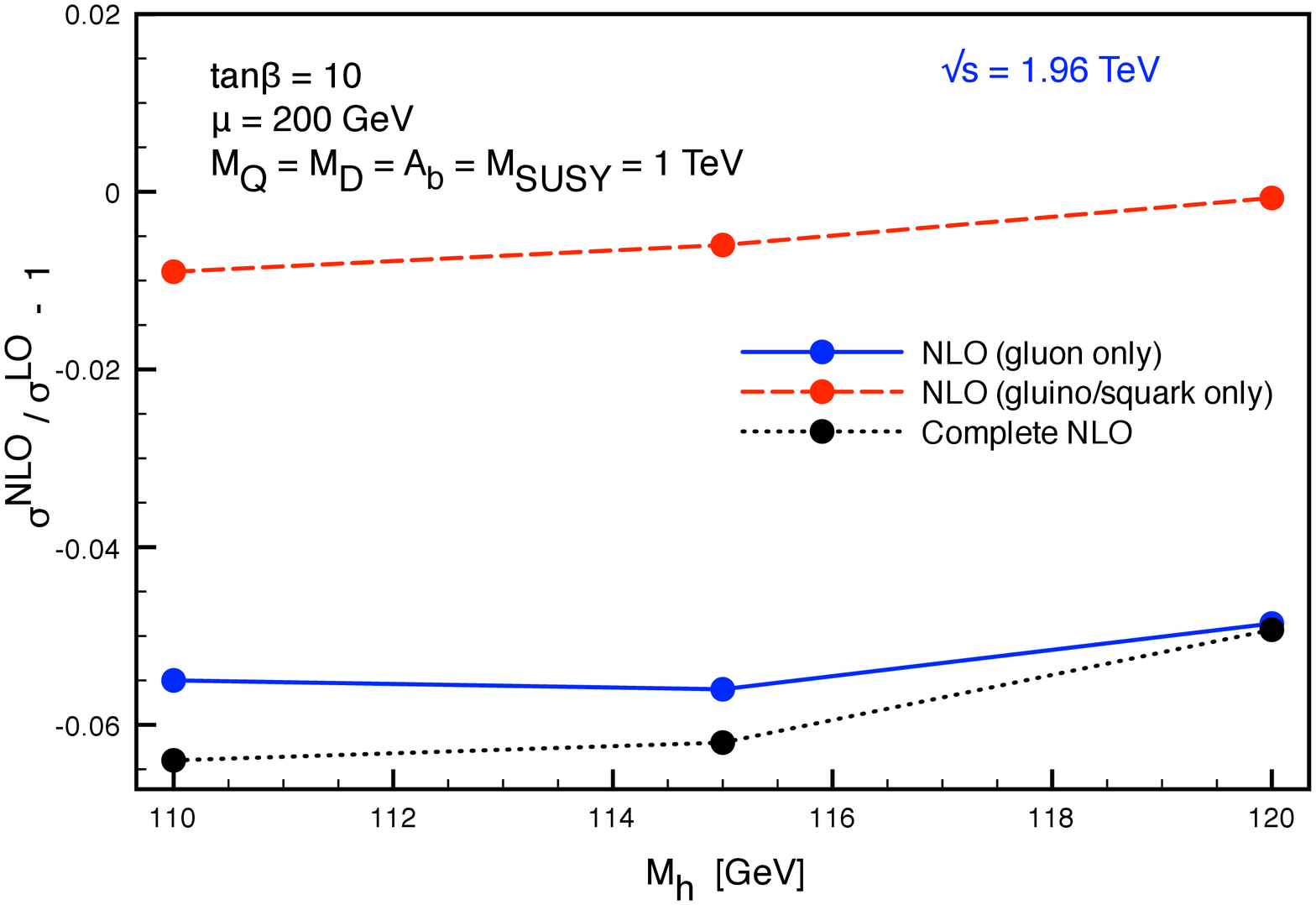}& 
\includegraphics[bb=8 24 700 700,scale=0.35]
{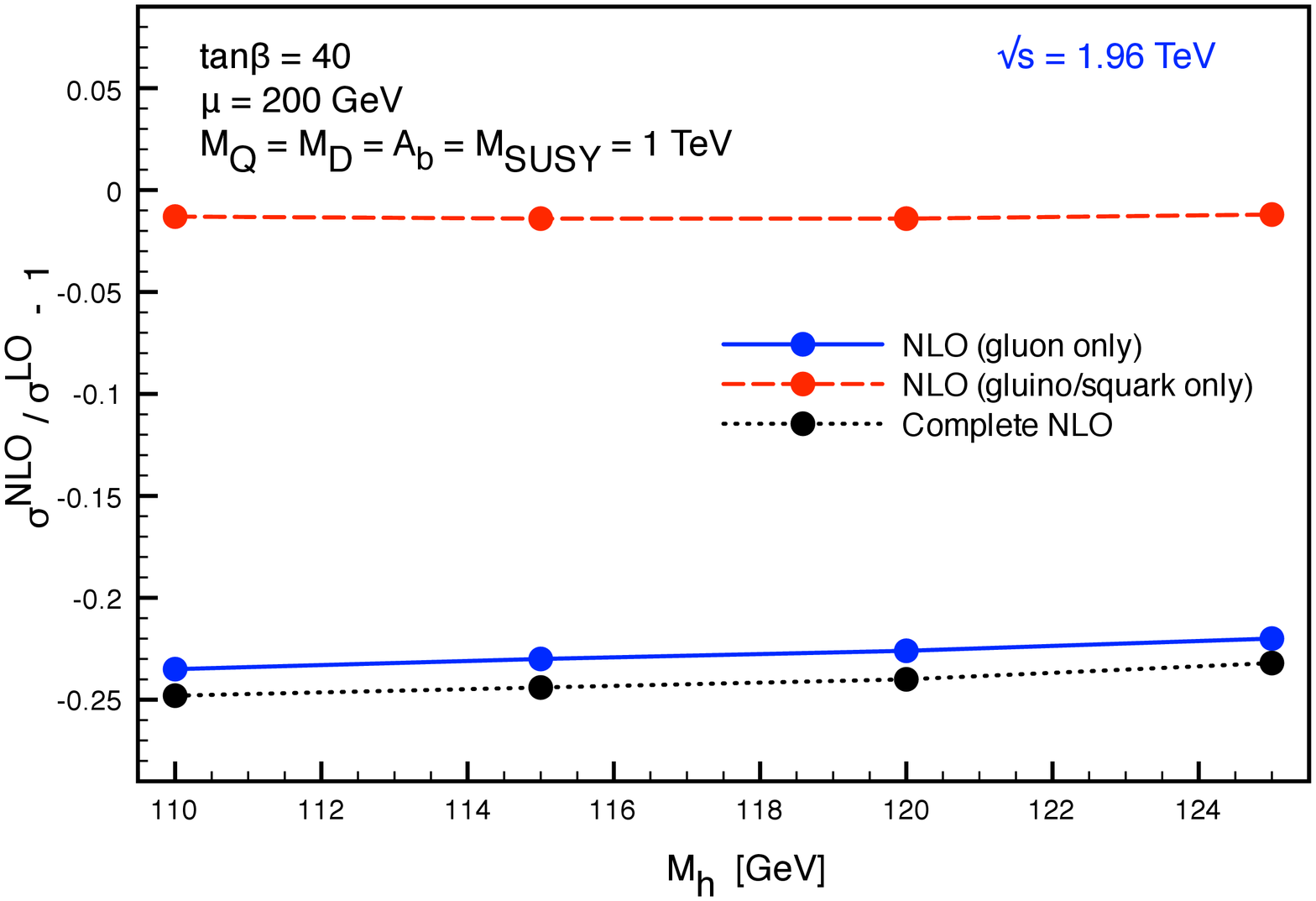}
\end{tabular}
\vspace*{8pt}
\caption[]{
Effects of NLO QCD and SQCD contributions on the rate for $p p
\rightarrow b h^0$ at the LHC.  The outgoing $b$ quark satisfies $p_T>20~GeV$
and $\mid \eta \mid <2$.  
The  NLO results use NLO PDFs, $2-$loop evolution
of $\alpha_s(\mu_R)$ and $\overline{m_b}(\mu_R)$, 
and the Yukawa couplings of Eq. \ref{mbdef}.
The plots are normalized to
the lowest order cross section of Eq. \ref{siglo}, which is computed with lowest order PDFs, 
$1$-loop evolution
of $\alpha_s(\mu_R)$ and $\overline{m_b}(\mu_R)$, and the lowest order
Yukawa couplings, $g_{b{\overline b}h}^{LO}$.}
\label{fg:Mhlight}
\end{center}
\end{figure}

\begin{figure}[hbtp!]
\begin{center}
\hskip -.2in
\includegraphics[bb=8 24 700 700,scale=0.35]
{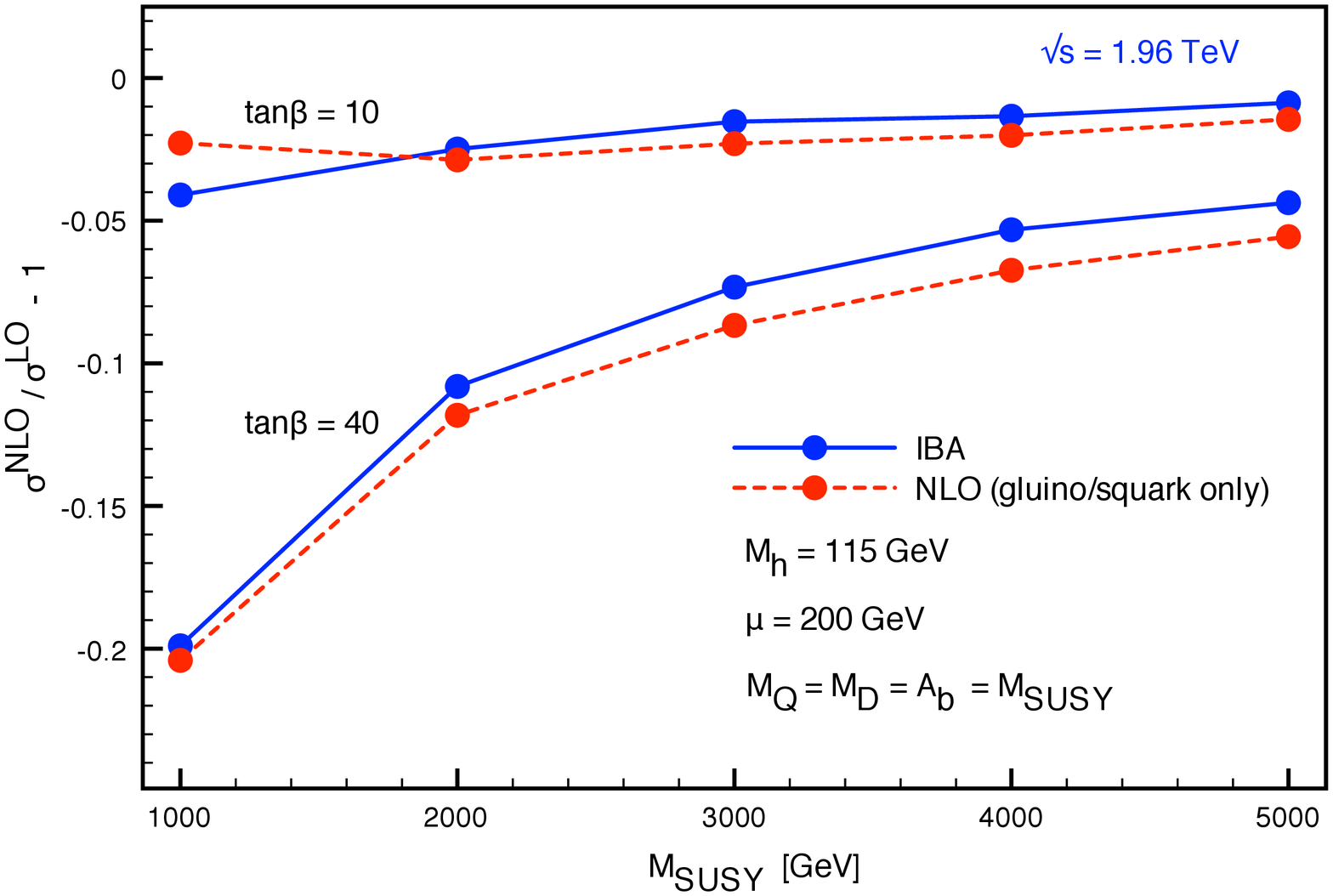}  \hskip .25in
\vspace*{8pt}
\caption[]{Comparison of the full SQCD calculation (Eq. \ref{sqcdsig}) 
with the 
Improved Born Approximation (Eq. \ref{sigiba}) on the rate for $p {\overline p}
\rightarrow b h^0$ at the Tevatron.  The outgoing $b$ quark satisfies $p_T>20~GeV$
and $\mid \eta \mid <2$.  }
\label{fg:Mhsusy}
\end{center}
\end{figure}

Results for the lightest MSSM scalar, $h^0$,
are shown in Figs. \ref{fg:Mhlight} and \ref{fg:Mhsusy} for the 
Tevatron.\footnote{A preliminary study of the SQCD contributions 
to $bh^0$ production in the MSSM 
appears in Ref. \cite{Cao:2002ja}.  This work used a
renormalization framework in which the heavy squarks and gluino do
not decouple and hence their results are not directly comparable
to ours. We find, however, qualitative agreement with their results.}
Fig. \ref{fg:Mhlight} shows the dependence of the 
complete set of SQCD contributions
(Eq. \ref{sqcdsig}) and the pure NLO QCD 
result (Eq. \ref{qcdsig})  on the lightest
Higgs mass for $\tan\beta=10$ and $40$. The curve marked
``gluon (only)'' (Eq. \ref{qcdsig}) 
includes the NLO QCD corrections and uses the
effective Yukawa couplings of Eq. \ref{mbdef}, while the SQCD
curve (Eq. \ref{sqcdsig})
incorporates the full set of SQCD corrections (box diagrams, etc).
 We see that the effects of 
squark and gluino loops which are not absorbed into the
the effective Yukawa couplings  are typically less than $1-2\%$ 
for both
$\tan \beta=10$ and $\tan\beta=40$ and can safely be neglected.
We also note that the sizes of the NLO corrections (both pure QCD
and SQCD) do not have  strong dependences on the mass of the
produced Higgs boson.

Fig. \ref{fg:Mhsusy} compares the  SQCD result (Eq. 
\ref{sqcdsig}) with that
obtained in the Improved Born Approximation (IBA), Eq. \ref{sigiba}.  For both values of $\tan\beta$ plotted,
the Improved Born Approximation  is an excellent approximation.  
This is an important result of our calculation because it demonstrates that
the effective Lagrangian approach is extremely accurate for $b\phi$ production.
It is apparent that the squark and gluino  contributions which are not
included in the IBA approximation give a negligible contribution to the rate.
Figure \ref{fg:Mhsusy} also exhibits the slow approach to the decoupling limit already
noted in Ref. \cite{Haber:2000kq}.  The decoupling limit, where the SQCD corrections
do not contribute to the rate, corresponds to
 $\sigma_{NLO}/\sigma_{LO}-1=0$ in this plot.  The approach to the decoupling limit
is significantly slower for large $\tan\beta$ than for small $\tan\beta$, as is apparent
in Eq. \ref{decouple}. Even for $M_{SUSY}$ as large as $5$~TeV, the SQCD
effects present in the couplings of the  effective Lagrangian of 
Eq. \ref{mbdef} are still ${\cal O}(5\%)$ for $\tan\beta=40$.  

\begin{figure}[hbtp!]
\begin{center}
\begin{tabular}{rl}
\includegraphics[bb=8 24 700 700,scale=0.35]
{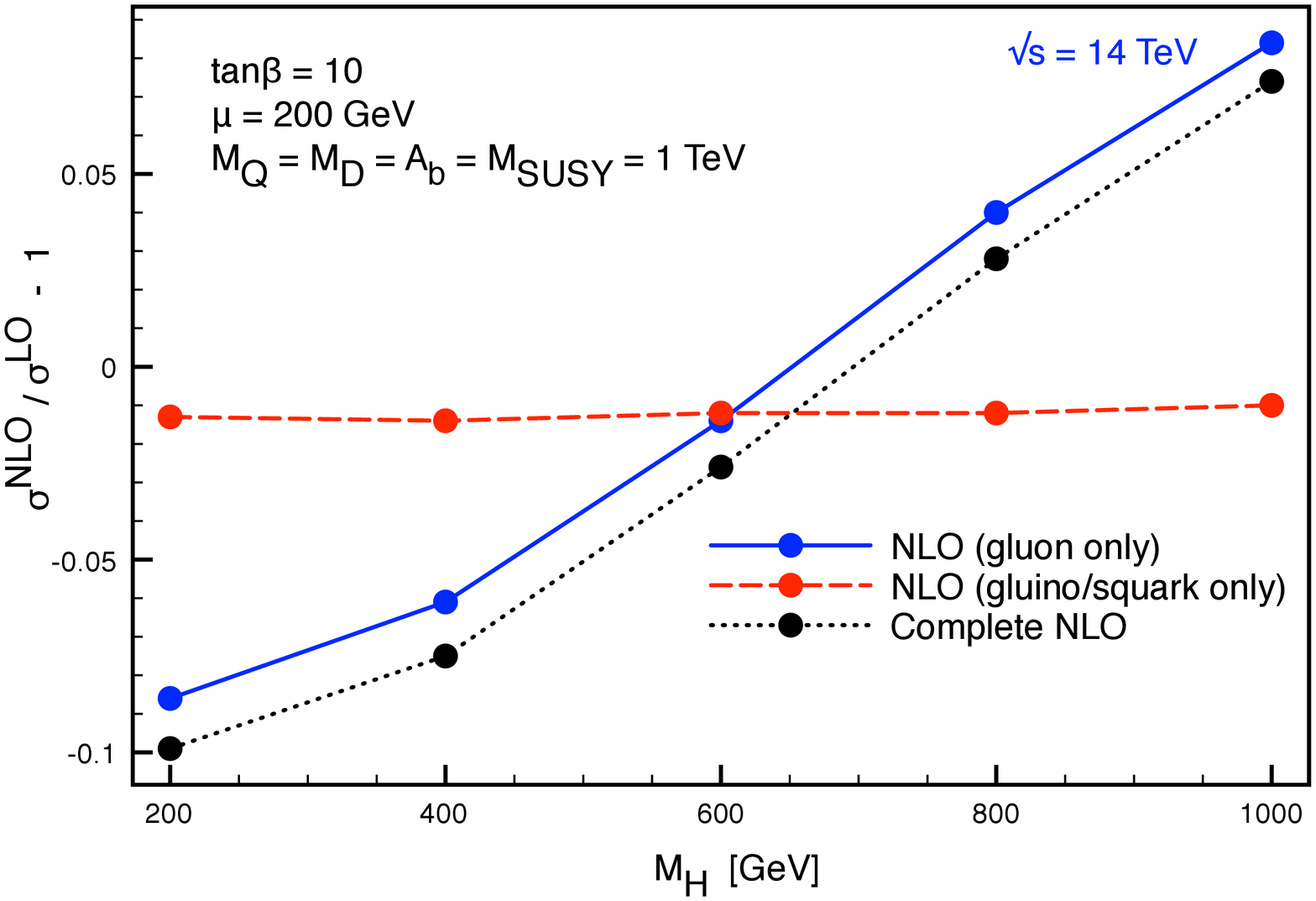} & 
\includegraphics[bb=8 24  700 700,scale=0.35]
{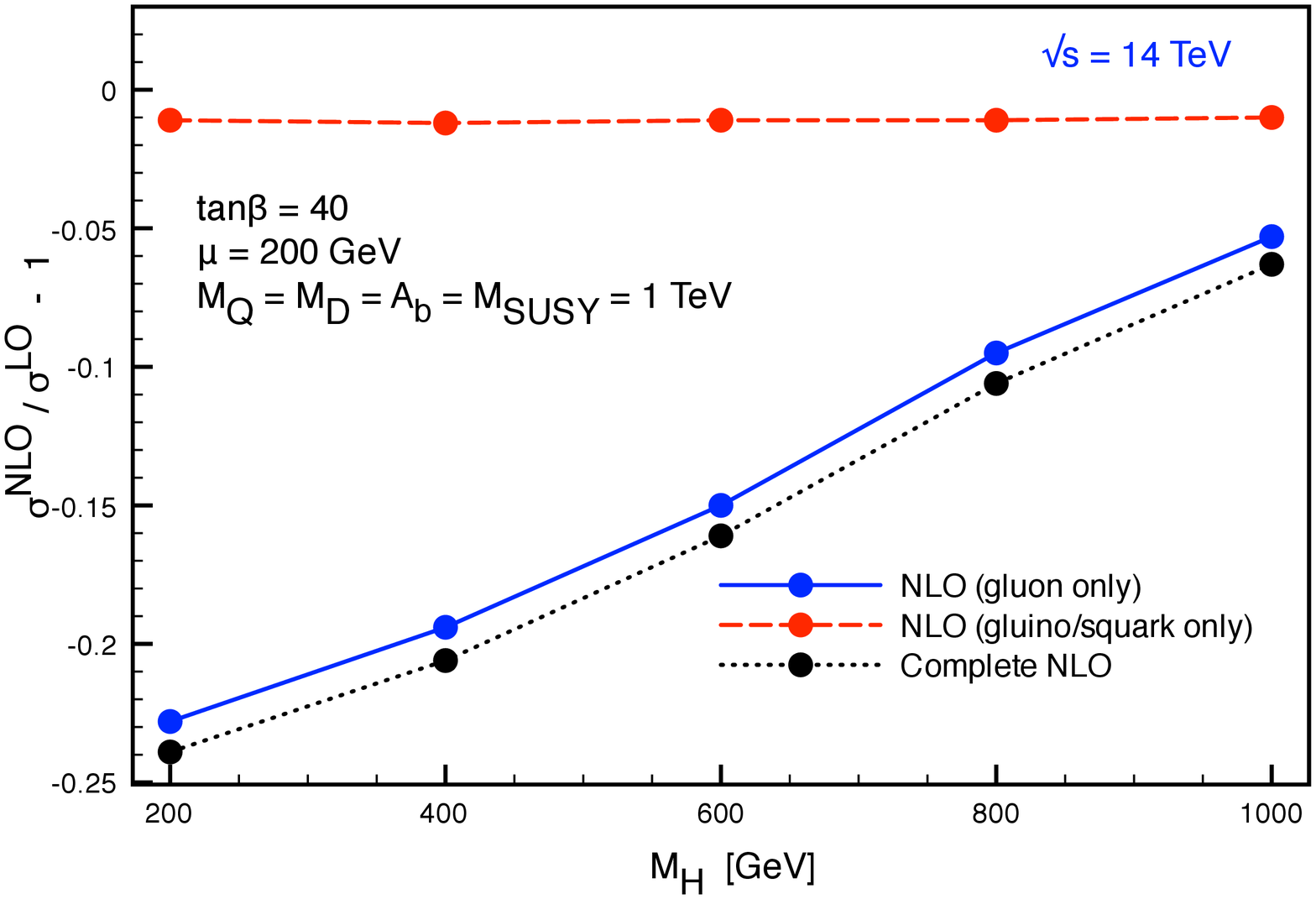}
\end{tabular}
\vspace*{8pt}
\caption[]{
Effects of NLO QCD and SQCD contributions on the rate for $p p
\rightarrow b H^0$ at the LHC.  The outgoing $b$ quark satisfies $p_T>20~GeV$
and $\mid \eta \mid <2.5$.  
The  NLO results use NLO PDFs, $2-$loop evolution
of $\alpha_s(\mu_R)$ and $\overline{m_b}(\mu_R)$, 
and the Yukawa couplings of Eq. \ref{mbdef}.
  The plots are normalized to
the lowest order cross section of Eq. \ref{siglo}, which is computed with lowest order PDFs, 
$1$-loop evolution
of $\alpha_s(\mu_R)$ and $\overline{m_b}(\mu_R)$, and the lowest order
Yukawa couplings, $g_{b{\overline b}h}^{LO}$.}
\label{fg:MHtanb}
\end{center}
\end{figure}

\begin{figure}[hbtp!]
\begin{center}
\includegraphics[bb=8 24 700 700,scale=0.35]
{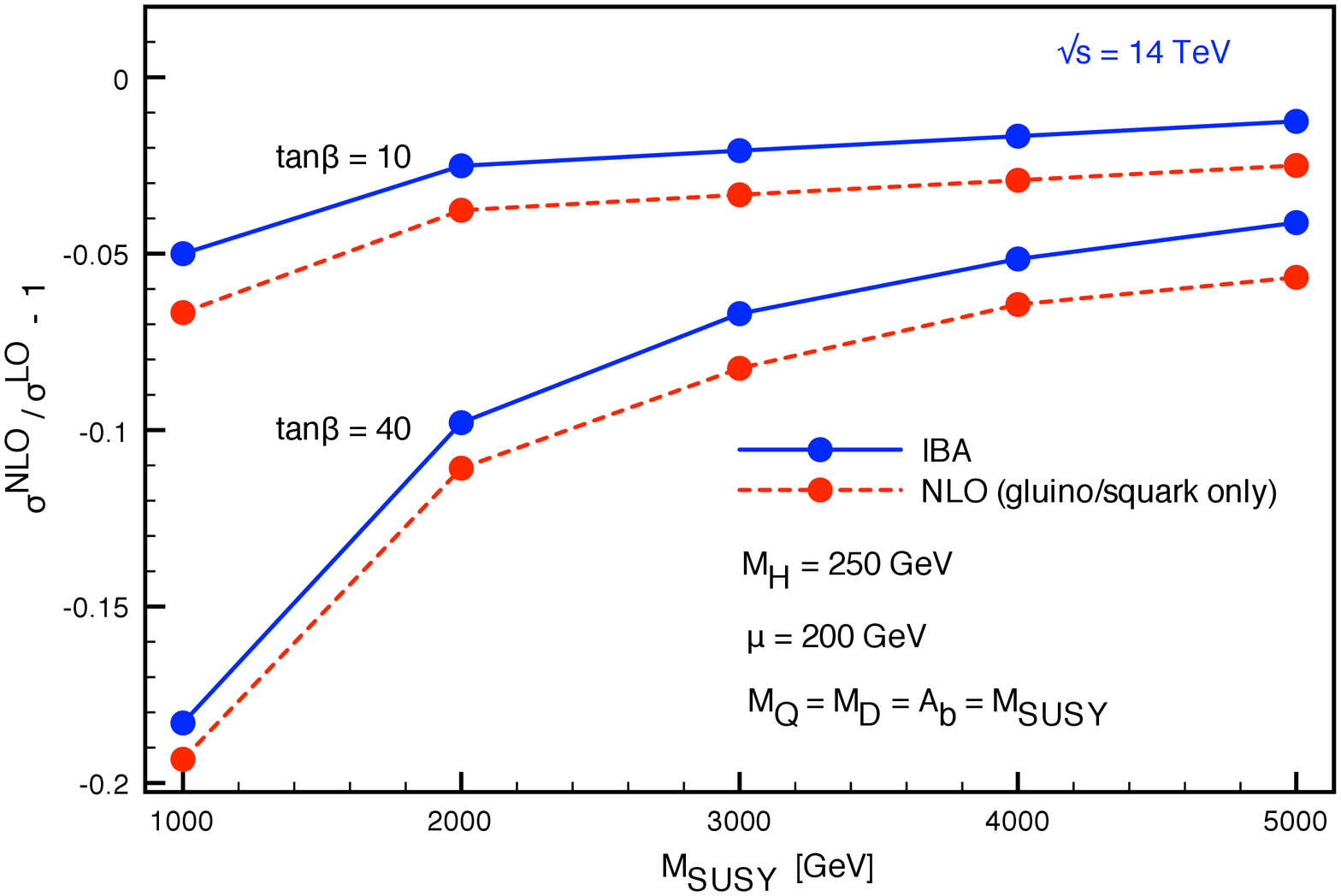} 
\vspace*{8pt}
\caption[]{
Comparison of the full SQCD calculation (Eq. \ref{sqcdsig})
 with the Improved Born Approximation (Eq. \ref{sigiba})
for the rate for $p p
\rightarrow b H^0$ at the LHC.  The outgoing $b$ quark satisfies $p_T>20~GeV$
and $\mid \eta \mid <2.5$. } 
\label{fg:MHsusy}
\end{center}
\end{figure}

Results for the heavier neutral MSSM scalar, $H^0$,
are shown in Figs. \ref{fg:MHtanb} and \ref{fg:MHsusy} for the LHC.  
Fig. \ref{fg:MHtanb} shows the dependence of the SQCD corrections on the heavy
Higgs mass for $\tan\beta=10$ and $40$.   The 
NLO pure QCD results (Eq. \ref{qcdsig})
 are quite sensitive to $M_H$, while the SQCD corrections (Eq. \ref{sqcdsig})
are relatively independent of the Higgs mass and quite small.
The complete NLO rate including all QCD and SQCD effects (Eq. \ref{comdef})
differs by less than $1~\%$ from the ``NLO (gluon only)'') calculation
where the SQCD effects are included only via the Yukawa couplings of 
the effective Lagrangian.
Fig. \ref{fg:MHsusy} compares the full SQCD result with that
obtained in the Improved Born Approximation (IBA) and as is the case for the lighter
Higgs boson, the
Improved Born Approximation closely approximates the full SQCD
calculation.  The slow approach to the decoupling limit for the $H^0$ is similar
to that seen for the $h^0$.

\begin{figure}[hbtp!]
\begin{center}
\begin{tabular}{rl}
\includegraphics[bb=8 24 700 700,scale=0.35]
{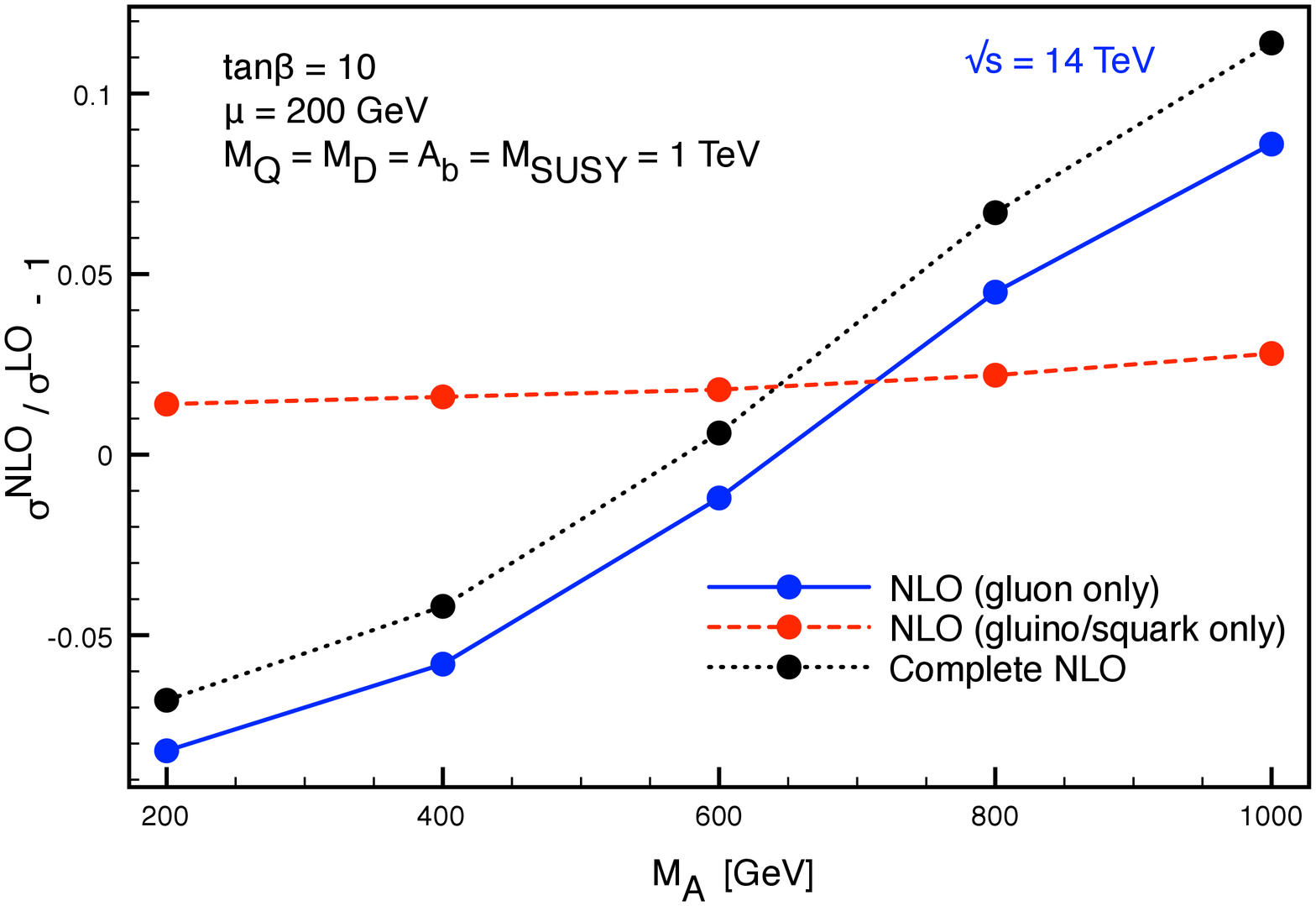} & 
\includegraphics[bb=8 24  700 700,scale=0.35]
{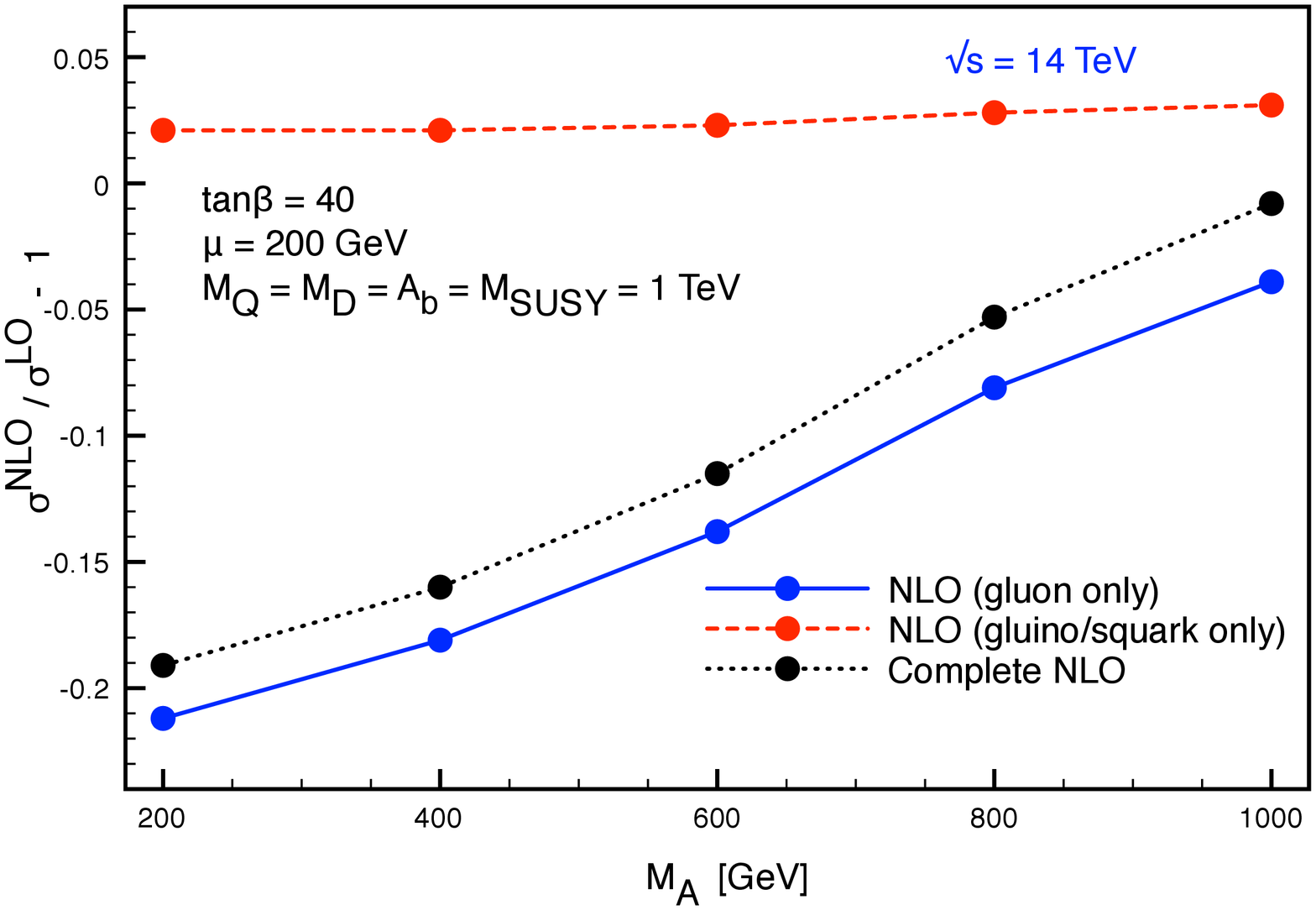}
\end{tabular}
\vspace*{8pt}
\caption[]{
Effects of NLO QCD and SQCD contributions on the rate for $p p
\rightarrow b A^0$ at the LHC.  The outgoing $b$ quark satisfies $p_T>20~GeV$
and $\mid \eta \mid <2.5$.  
The  NLO results use NLO PDFs, $2-$loop evolution
of $\alpha_s(\mu_R)$ and $\overline{m_b}(\mu_R)$, 
and the Yukawa couplings of Eq. \ref{mbdef}.
  The plots are normalized to
the lowest order cross section of Eq. \ref{siglo}, which is computed with lowest order PDFs, 
$1$-loop evolution
of $\alpha_s(\mu_R)$ and $\overline{m_b}(\mu_R)$, and the lowest order
Yukawa couplings, $g_{b{\overline b}h}^{LO}$.}
\label{fg:MAtanb}
\end{center}
\end{figure}

\begin{figure}[hbtp!]
\begin{center}
\includegraphics[bb=8 24 700 700,scale=0.35]
{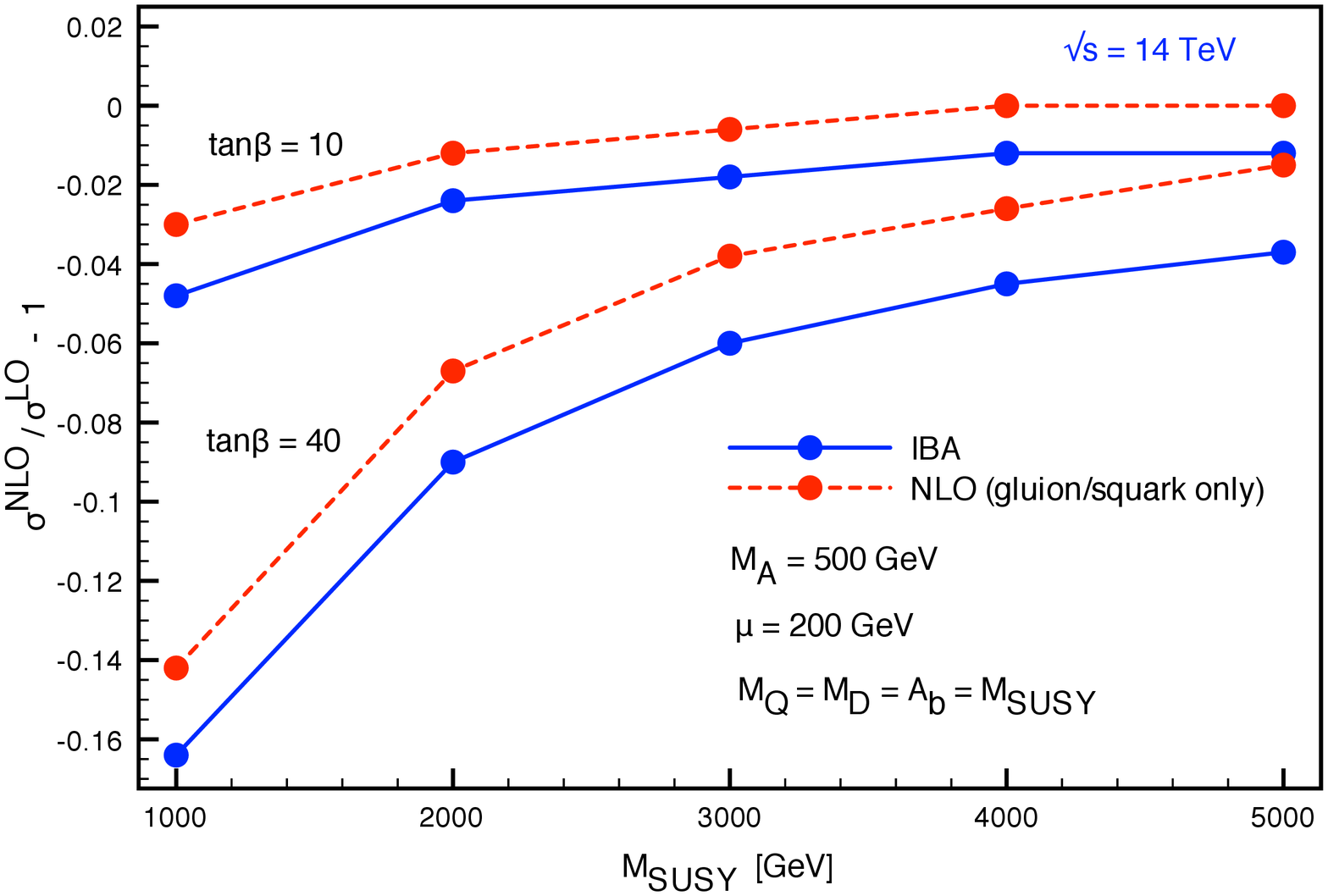} 
\vspace*{8pt}
\caption[]{Comparison of the full SQCD calculation (Eq. \ref{sqcdsig}) with the
Improved Born Approximation (Eq. \ref{sigiba})
for
$p p
\rightarrow b A^0$ at the LHC.  The outgoing $b$ quark satisfies $p_T>20~GeV$
and $\mid \eta \mid <2.5$.  }
\label{fg:MAsusy}
\end{center}
\end{figure}

Results for the MSSM pseudoscalar, $A^0$,
are shown in Figs. \ref{fg:MAtanb} and \ref{fg:MAsusy} for the LHC.
Fig. \ref{fg:MAtanb} shows the dependence of the SQCD corrections on the pseudoscalar
mass for $\tan\beta=10$ and $40$.  The SQCD effects (Eq. \ref{sqcdsig})
 are positive
and are of a similar magnitude as those found for $H^0$ production and again
are relatively independent of the Higgs mass, $M_A$.  
Fig. \ref{fg:MAsusy}
 compares the full SQCD result with that obtained in the Improved
Born Approximation.  In this case, the Improved Born Approximation
gives a prediction which differs from 
the full SQCD rate by about $-3\%$ for both $\tan\beta=10$ and $40$ 
and we again 
see the slow approach to the decoupling limit.  This plot fixes $M_A=500~GeV$
and varies $M_{SUSY}$.  From Eq. \ref{decma}, 
it is clear that for decoupling to occur
we need to also take $M_A\rightarrow \infty$.  For fixed $M_A$, even for
$M_{SUSY}=5~TeV$, the effect of squark and gluino loops are still ${\cal O}
(5-10)\%$.

\section{Conclusions}

\begin{figure}[t]
\begin{center}
\includegraphics[scale=0.4]{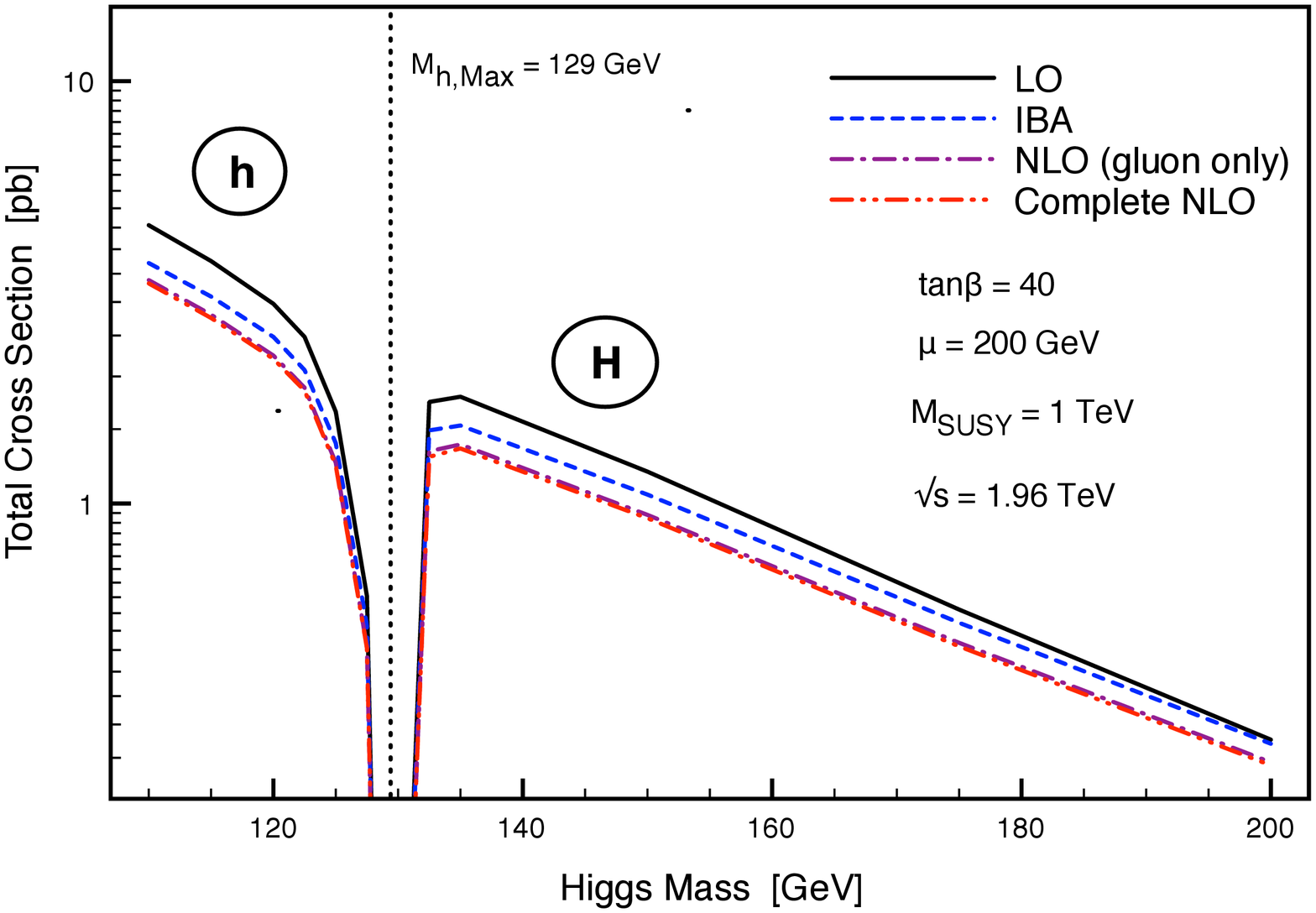} 
\caption[]{Complete NLO ${\cal O}(\alpha_s^2)$ results for $
p {\overline p}\rightarrow b h^0 (H^0)$ 
at the Tevatron.}
\label{fg:tevfinal}
\end{center}
\end{figure}

\begin{figure}[t]
\begin{center}
\includegraphics[scale=0.4]{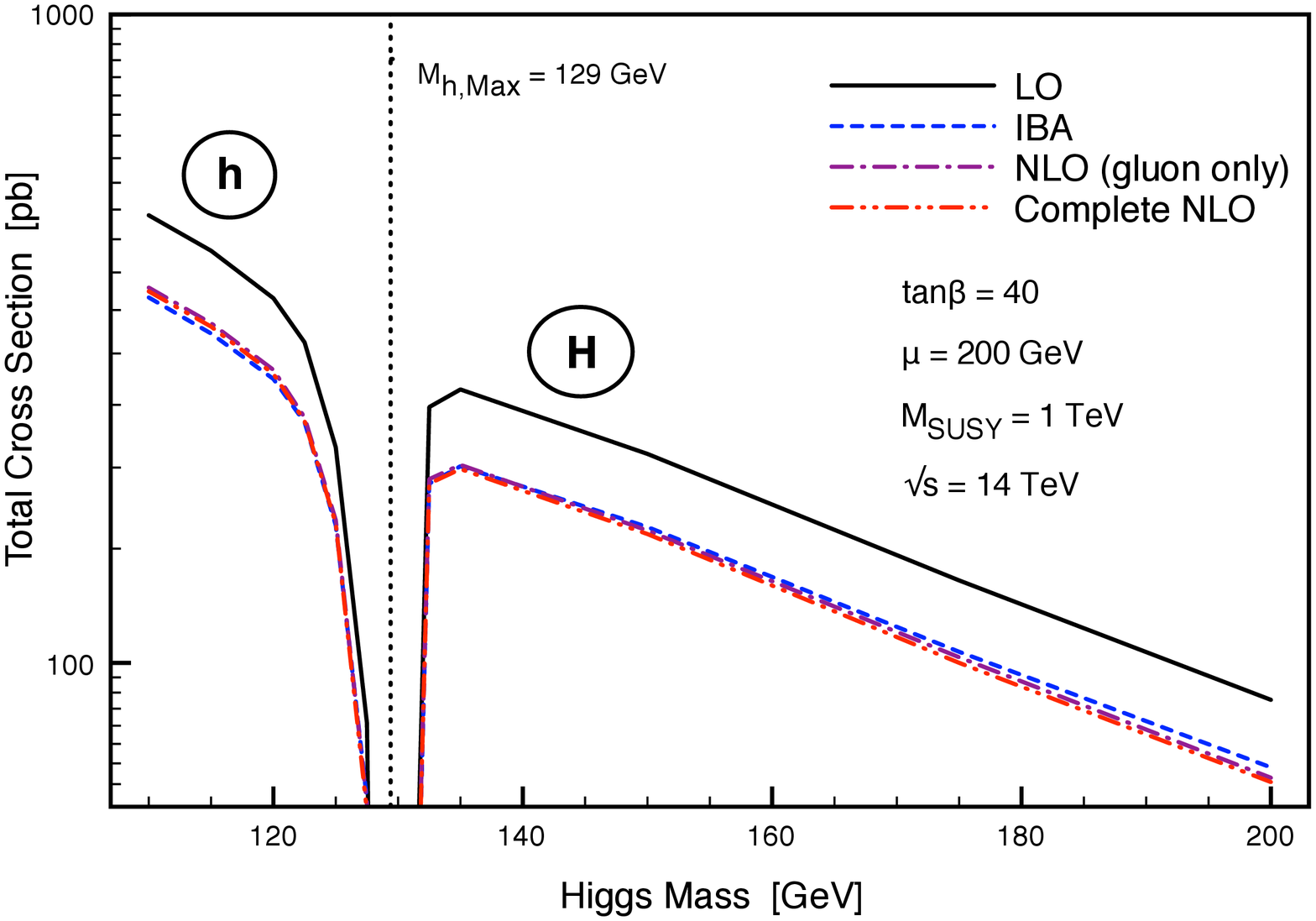} 
\caption[]{Complete NLO ${\cal O}(\alpha_s^2)$ results
for $
p p\rightarrow b h^0 (H^0)$ 
at the LHC.} 
\label{fg:lhcfinal}
\end{center}
\end{figure}

We have computed the ${\cal O}(\alpha_s^2)$ SUSY QCD corrections from squark
and gluino loops for the associated production of an MSSM Higgs boson and a 
$b$ quark.  When the Yukawa couplings are normalized with the
effective Lagrangian of Eq. \ref{mbdef}, the remaining SQCD corrections 
from squark
and gluino loops  are typically of order a few percent.
We have therefore explicitly demonstrated
that the effective Lagrangian approach to calculating the effects of
squark and gluino loops is 
extremely accurate and can be reliably used to approximate the $b\phi$ cross
sections. 
In Figs.~\ref{fg:tevfinal} and \ref{fg:lhcfinal}, we summarize our results by showing
 the complete NLO predictions for
$b$ Higgs associated production at the Tevatron and the 
LHC.\footnote{The $bh$ cross section decreases rapidly as the 
maximum allowed value of
$M_h$ is approached from below due to the suppression of the $b{\overline b}h$
coupling.  Similarly, the $bH$ cross section decreases as $M_{h,{Max}}$ is
approached from above due to the suppression in this region 
of the $ b {\overline b}H$ coupling.}  
These figures include all NLO ${\cal O}(\alpha_s^2)$ contributions and represent
the most complete calculation available. 

\begin{acknowledgments}

The authors would like to thank C. Kao and Y. Wang
for useful discussions.
This work is supported by the U.S. Department of
Energy under grant
DE-AC02-98CH10886.
S.D. thanks the SLAC theory group for their hospitality, where
this work was completed.

\end{acknowledgments}
\appendix
\section{Scalar Integrals and Tensor Coefficients}
\label{app:P-V}

The scalar integrals are defined as:
\begin{eqnarray}
\label{eq:A0}
{i\over 16\pi^2}A_0(M_0) &=& \int \frac{d^nk}{(2\pi)^n} \frac{1}{N_0}\,, 
\nonumber\\
\label{eq:B0}
{i\over 16\pi^2}B_0(p_1^2;M_0,M_1) &=& \int \frac{d^nk}{(2\pi)^n} \frac{1}{N_0 N_1}\,,
\nonumber\\
\label{eq:C0}
{i\over 16\pi^2}C_0(p_1^2,p_2^2;M_0,M_1,M_2) &=& \int \frac{d^nk}{(2\pi)^n} \frac{1}{N_0 N_1 N_2}\,, 
\nonumber\\
\label{eq:D0}
{i\over 16\pi^2}D_0(p_1^2,p_2^2,p_3^2;M_0,M_1,M_2,M_3) &=&
 \int \frac{d^nk}{(2\pi)^n} \frac{1}{N_0 N_1 N_2 N_3}\,, 
\end{eqnarray}
where:
\begin{eqnarray}
N_0 &=& k^2 - M_0^2 
\nonumber\\
N_1 &=& (k + p_1)^2 - M_1^2 
\nonumber\\
N_2 &=& (k + p_1 + p_2)^2 - M_2^2 
\nonumber\\
N_3 &=& (k + p_1 + p_2 + p_3)^2 - M_3^2 \,.
\end{eqnarray}

The tensor integrals encountered are expanded in terms of the 
external momenta $p_i$ and the metric tensor $g^{\mu\nu}$.  For the two-point
function we write:
\begin{eqnarray}
{i\over 16\pi^2}B^\mu(p_1^2;M_0,M_1) &=& \int \frac{d^nk}{(2\pi)^n} \frac{k^\mu}{N_0 N_1} 
\nonumber\\
&\equiv& {i\over 16\pi^2}p_1^\mu B_1(p_1^2,M_0,M_1)\,,
\end{eqnarray}
while for the three-point functions we have both rank-one and rank-two tensor 
integrals which we expand as:
\begin{eqnarray}
C^\mu(p_1^2,p_2^2;M_0,M_1,M_2) 
&=& p_1^\mu C_{11} + p_2^\mu C_{12} \,,
\nonumber\\
C^{\mu\nu}(p_1^2,p_2^2;M_0,M_1,M_2) &=& p_1^\mu p_1^\nu C_{21} + 
  p_2^\mu p_2^\nu C_{22} \nonumber\\
&+& (p_1^\mu p_2^\nu + p_1^\nu p_2^\mu) C_{23} + g^{\mu\nu} C_{24} \,,
\end{eqnarray} 
where:
\begin{equation}
{i\over 16\pi^2}
C^\mu (C^{\mu\nu})(p_1^2,p_2^2;M_0,M_1,M_2)  \equiv
\int \frac{d^nk}{(2\pi)^n} \frac{k^\mu (k^\mu k^\nu)}{N_0 N_1 N_2}
\end{equation}

Finally, for the box diagrams, we encounter only rank-one tensor integrals which
are written in terms of the Passarino-Veltmann coefficients as:
\begin{eqnarray}
{i\over 16\pi^2}D^\mu(p_1^2,p_2^2;M_0,M_1,M_2)  &\equiv&
\int \frac{d^nk}{(2\pi)^n} \frac{k^\mu}{N_0 N_1 N_2 N_3} 
\nonumber\\
&=& {i\over 16\pi^2}
\biggl\{ p_1^\mu D_{11} + p_2^\mu D_{12} + p_3^\mu D_{13}\biggr\} \,.
\end{eqnarray}

\section{One Loop SQCD Coefficients}

 The cofficients for each diagram are given below in the notation of Eq. \ref{eq:ampNLO}.

\begin{figure}[hbtp!]
\begin{center}
\begin{tabular}{lr}
\includegraphics[scale=0.7]
{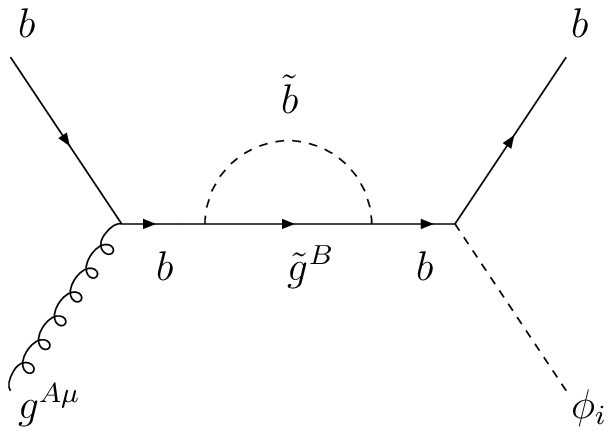} & 
\includegraphics[scale=0.7]
{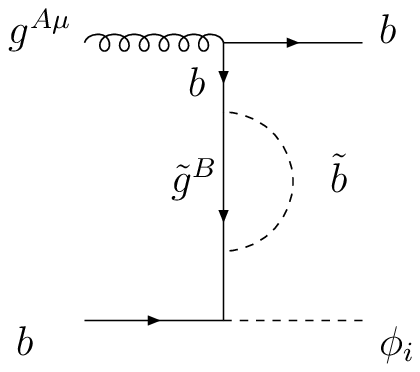}  
\end{tabular}
\vspace*{8pt}
\caption[]{Self- energy diagrams, $S_1$ and $S_2$.}
\label{fg:selffeyn}
\end{center}
\end{figure}

The self-energy contributions are shown in Fig. \ref{fg:selffeyn}.

{\bf{\underline{Diagram $S_1$:}}}

\begin{eqnarray}
X_{S_1}^{(s)} &=&X_{S_1}^{(1)}= 0 
\nonumber\\
X_{S_1}^{(t)} &=& -\frac{4}{3} g_{b\bar{b}\phi} \biggl[\frac{B_1(t;m_{\tilde{g}},m_{\tilde{b_1}})
  + B_1(t;m_{\tilde{g}},m_{\tilde{b_2}})}{t} \biggr]
 \,.
\end{eqnarray}

{\bf{\underline{Diagram $S_2$:}}}

\begin{eqnarray}
X_{S_2}^{(s)} &=& -\frac{4}{3} g_{b\bar{b}\phi} \biggl[\frac{B_1(s;m_{\tilde{g}},m_{\tilde{b_1}})
  + B_1(s;m_{\tilde{g}},m_{\tilde{b_2}})}{s} \biggr]
\nonumber\\
X_{S_2}^{(t)} &=& X_{S_2}^{(1)}=0\, .
\end{eqnarray}

\begin{figure}[hbtp!]
\begin{center}
\begin{tabular}{lr}
\includegraphics[scale=0.7]
{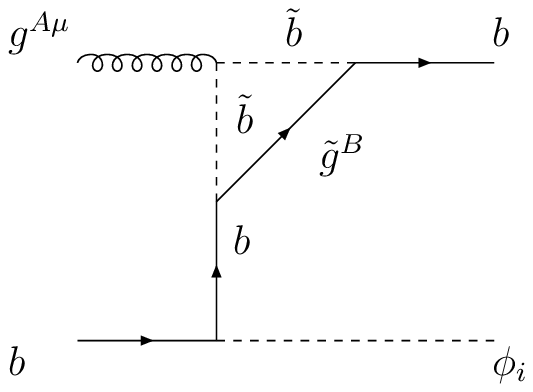} & 
\includegraphics[scale=0.7]
{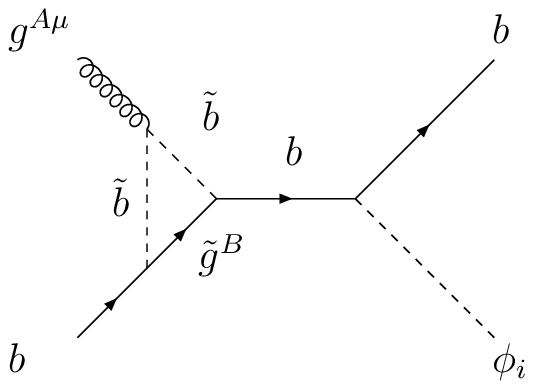}  
\end{tabular}
\vspace*{8pt}
\caption[]{Vertex diagrams, $V_1$ and $V_2$.}
\label{fg:virt1feyn}
\end{center}
\end{figure}

\begin{figure}[hbtp!]
\begin{center}
\begin{tabular}{lr}
\includegraphics[scale=0.7]
{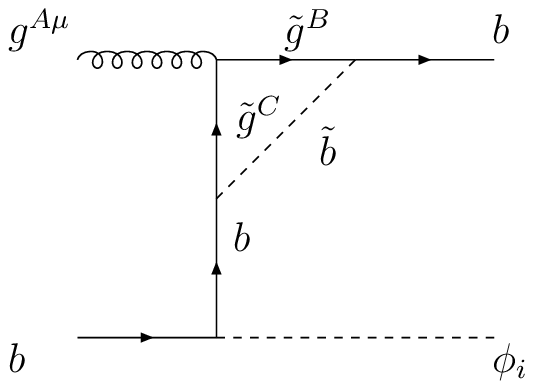} & 
\includegraphics[scale=0.7]
{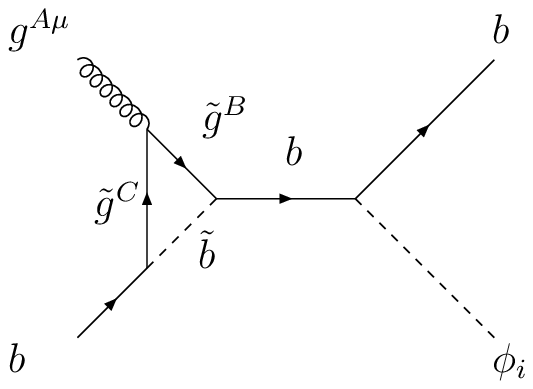}  
\end{tabular}
\vspace*{8pt}
\caption[]{Vertex diagrams, $V_3$ and $V_4$.}
\label{fg:virt2feyn}
\end{center}
\end{figure}

\begin{figure}[hbtp!]
\begin{center}
\begin{tabular}{lr}
\includegraphics[scale=0.7]
{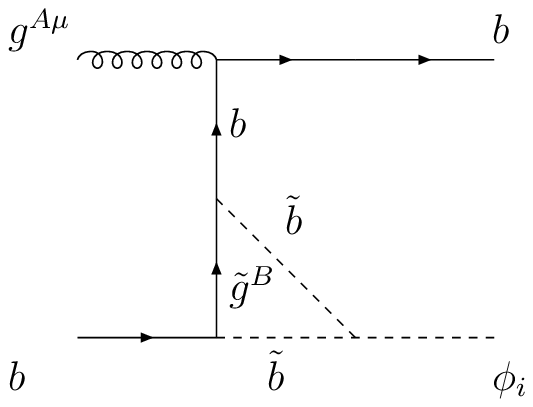} & 
\includegraphics[scale=0.7]
{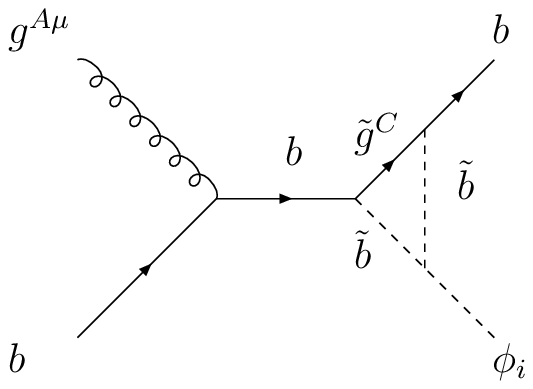}  
\end{tabular}
\vspace*{8pt}
\caption[]{Vertex diagrams, $V_5$ and $V_6$.}
\label{fg:virt3feyn}
\end{center}
\end{figure}

The virtual diagrams are shown in Figs. \ref{fg:virt1feyn}, \ref{fg:virt2feyn}
and \ref{fg:virt3feyn}.  An effective $b{\overline b} g$ vertex can be extracted 
from Figs. \ref{fg:virt1feyn} and \ref{fg:virt2feyn} and in the limit
where the $b$ quarks of the effective vertex are on-shell, our results
agree with Ref. \cite{Berge:2007dz}. An effective $b {\overline b}\phi_i$ vertex
can be found from Figs. \ref{fg:virt3feyn} and agrees with the results of 
Refs. \cite{Haber:2000kq,Hall:1993gn,Carena:1999py} when the $b$ quarks are on shell.

{\bf{\underline{Diagram $V_1$ :}}}

\begin{eqnarray}
X_{V_1}^{(s)} &=&  -\frac{g_{b\bar{b}\phi}}{6} \sum_{i = 1}^2   
  \biggl[ C_{12} + C_{23} \biggr](0,0; m_{\tilde{g}},m_{\tilde{b_i}},
  m_{\tilde{b_i}}) 
\nonumber\\
X_{V_1}^{(t)} &=&   \frac{g_{b\bar{b}\phi}}{6} \sum_{i = 1}^2  
  \biggl[ \frac{2}{t} C_{24} + C_{12} + C_{23} \biggr] (0,0; m_{\tilde{g}},m_{\tilde{b_i}},
  m_{\tilde{b_i}}) 
\nonumber\\
X_{V_1}^{(1)} &=&   \frac{g_{b\bar{b}\phi}}{3} \sum_{i = 1}^2   
  \biggl[ C_{12} + C_{23} \biggr](0,0; m_{\tilde{g}},m_{\tilde{b_i}},
  m_{\tilde{b_i}})\, . 
\end{eqnarray}

{\bf{\underline{Diagram $V_2$ :}}}

\begin{eqnarray}
X_{V_2}^{(s)} &=& \frac{g_{b\bar{b}\phi}}{3} \sum_{i = 1}^2 \frac{C_{24}}
  {s}(s,0; m_{\tilde{g}},m_{\tilde{b_i}},m_{\tilde{b_i}}) 
\nonumber\\
X_{V_2}^{(t)} &=& 0 
\nonumber\\
X_{V_2}^{(1)} &=& \frac{g_{b\bar{b}\phi}}{3} \sum_{i = 1}^2   
  \biggl[ C_{11} - C_{12} + C_{21} - C_{23} \biggr]
  (s,0; m_{\tilde{g}},m_{\tilde{b_i}},m_{\tilde{b_i}})\, .
\end{eqnarray} 

\newpage
{\bf{\underline{Diagram $V_3$ :}}}

\begin{eqnarray}
X_{V_3}^{(s)} &=& -\frac{3}{2} g_{b\bar{b}\phi} \sum_{i = 1}^2   
  \biggl[ C_{12} + C_{23} \biggr](0,0; m_{\tilde{b_i}},
     m_{\tilde{g}},m_{\tilde{g}}) 
\nonumber\\
X_{V_3}^{(t)} &=& \frac{3}{2} g_{b\bar{b}\phi} \sum_{i = 1}^2   
  \biggl\{ \biggl[ \frac{1}{t} \biggl(2 C_{24} +
  (m_{\tilde{g}}^2 - m_{\tilde{b_i}}^2)C_0 \biggr) + C_{23}\biggr]
  (0,0; m_{\tilde{b_i}},m_{\tilde{g}},m_{\tilde{g}}) 
\nonumber\\
&& \,\,\,\,\,\,\,\,\,\,\,\,\,\,\,\,\,\,\,\,\,\, -
  \frac{B_0(0;m_{\tilde{g}},m_{\tilde{g}})}{t} \biggr\}   
\nonumber\\
X_{V_3}^{(1)} &=& 3 g_{b\bar{b}\phi} \sum_{i = 1}^2   
  \biggl[ C_{12} + C_{23}\biggr](0,0; m_{\tilde{b_i}},
     m_{\tilde{g}},m_{\tilde{g}})\, .
\end{eqnarray}

{\bf{\underline{Diagram $V_4$ :}}}

\begin{eqnarray}
X_{V_4}^{(s)} &=& \frac{3}{2} g_{b\bar{b}\phi} \sum_{i = 1}^2   
  \biggl\{ \biggl[ \frac{1}{s} \biggl(2 C_{24} +
  (m_{\tilde{g}}^2 - m_{\tilde{b_i}}^2)C_0 \biggr) - 2 C_{11}
  + C_{12}\biggr]
  (s,0; m_{\tilde{b_i}},m_{\tilde{g}},m_{\tilde{g}}) \nonumber\\
&& \,\,\,\,\,\,\,\,\,\,\,\,\,\,\,\,\,\,\,\,\,\, -
  \frac{B_0(0;m_{\tilde{g}},m_{\tilde{g}})}{s} \biggr\}   
\nonumber \\
X_{V_4}^{(t)} &=& 0 
\nonumber\\
X_{V_4}^{(1)} &=& 3 g_{b\bar{b}\phi} \sum_{i = 1}^2   
  \biggl[ C_{11} - C_{12} + C_{21} - C_{23} \biggr]
  (s,0; m_{\tilde{b_i}},m_{\tilde{g}},m_{\tilde{g}})\, .
\end{eqnarray}

{\bf{\underline{Diagram $V_5$ :}}}

\begin{eqnarray}
X_{V_5}^{(s)} &=&  X_{V_5}^{(1)}=0
\nonumber\\
X_{V_5}^{(t)} &=& \frac{4}{3} \biggl(\frac{m_{\tilde{g}}}{t}\biggr)
  \biggl\{ 2\cos{\tilde\theta}_b\sin{\tilde\theta}_b
  \biggl[ \tilde{C}_{i22} C_0(m_{\tilde{b_2}}^2, m_{\tilde{b_2}}^2) - 
          \tilde{C}_{i11} C_0(m_{\tilde{b_1}}^2, m_{\tilde{b_1}}^2) \biggr] \nonumber\\
&& +
  (\cos^2\tilde{\theta}_b - \sin^2{\tilde \theta}_b) \tilde{C}_{i12}
  \biggl[ C_0(m_{\tilde{b_1}}^2, m_{\tilde{b_2}}^2) + 
  C_0(m_{\tilde{b_2}}^2, m_{\tilde{b_1}}^2) \biggr] \biggr\}\, ,
\end{eqnarray} 
where the full arguments of the scalar integral and tensor coefficients are 
$(p_1^2,p_2^2; m_1^2,m_2^2,m_3^2) = (0,t;m_{\tilde{b_i}}^2,m_{\tilde{g}}^2,
m_{\tilde{b_j}}^2)$.

{\bf{\underline{Diagram $V_6$ :}}}

\begin{eqnarray}
X_{V_6}^{(s)} &=& \frac{4}{3} \biggl(\frac{m_{\tilde{g}}}{s}\biggr)
  \biggl\{ 2\cos\tilde{\theta}_b\sin\tilde{\theta}_b
  \biggl[ \tilde{C}_{i22} C_0(m_{\tilde{b_2}}^2, m_{\tilde{b_2}}^2) - 
          \tilde{C}_{i11} C_0(m_{\tilde{b_1}}^2, m_{\tilde{b_1}}^2) \biggr] \nonumber\\
&& +
  (\cos^2\tilde{\theta}_b - \sin^2\tilde{\theta}_b) \tilde{C}_{i12}
  \biggl[ C_0(m_{\tilde{b_1}}^2, m_{\tilde{b_2}}^2) + 
  C_0(m_{\tilde{b_2}}^2, m_{\tilde{b_1}}^2) \biggr] \biggr\}
\nonumber\\
X_{V_6}^{(t)} &=& X_{V_6}^{(1)}=0 \, ,
\end{eqnarray} 
where the full arguments of the scalar integral and tensor coefficients are 
$(p_1^2,p_2^2; m_1^2,m_2^2,m_3^2) = (0,s;m_{\tilde{b_i}}^2,
m_{\tilde{g}}^2,m_{\tilde{b_j}}^2)$.

\subsection{The Box Diagrams}

\begin{figure}[hbtp!]
\begin{center}
\begin{tabular}{lcr}
\includegraphics[scale=0.6]
{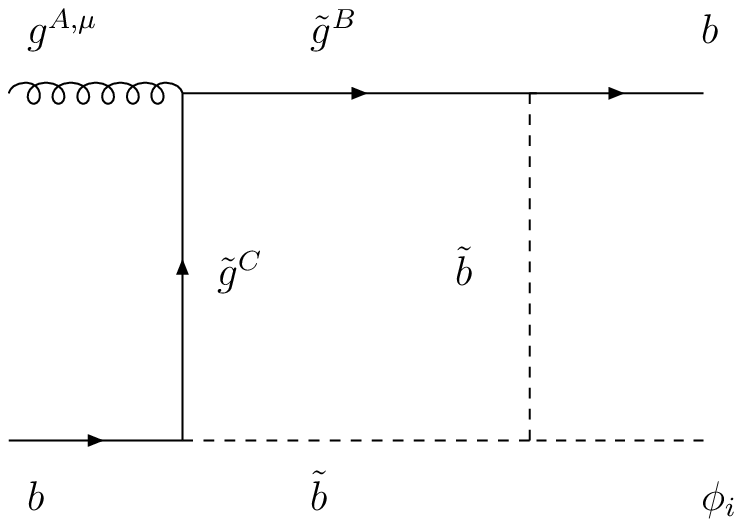} & 
\includegraphics[scale=0.6]
{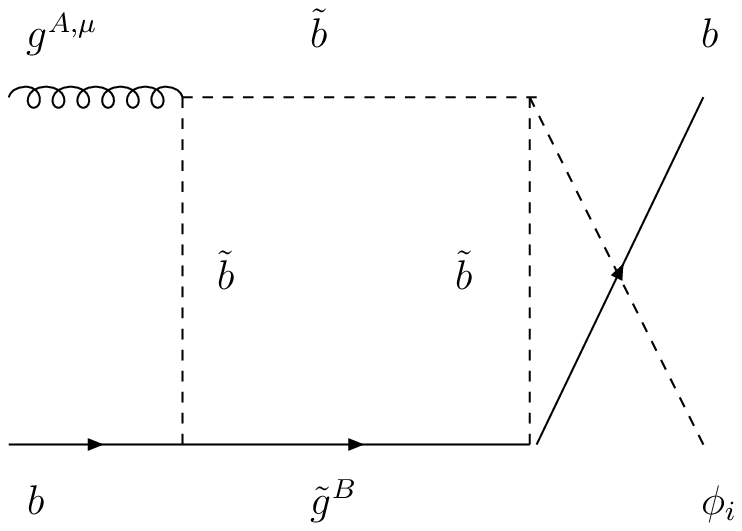} & 
\includegraphics[scale=0.6]
{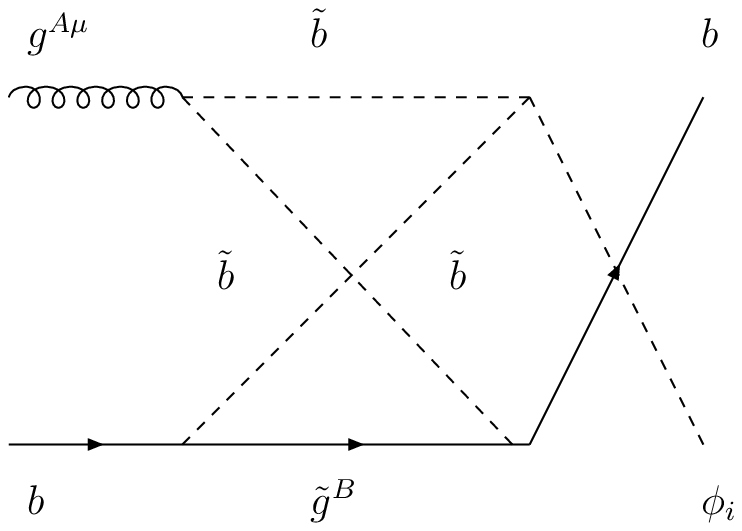} 
\end{tabular}
\vspace*{8pt}
\caption[]{Box diagrams $B_1$, $B_2$ and $B_3$.}
\label{fg:boxfeyn}
\end{center}
\end{figure}
The box diagrams are shown in Fig. \ref{fg:boxfeyn}.  The results of
Eqs. \ref{fg:box1}, \ref{fg:box2}
and
\ref{fg:box3} sum over the contributions of ${\tilde b}_1$ and ${\tilde b}_2$.

\newpage
{\bf{\underline{Diagram $B_1$:}}}

\begin{eqnarray}
X_{B_1}^{(s)} &=& \frac{3}{2} m_{\tilde{g}} 
  \biggl\{ \sin 2{\tilde\theta}_b
  \biggl[ \tilde{C}_{i22} (D_0 + D_{13})(m_{\tilde{b_2}}^2, m_{\tilde{b_2}}^2) - 
          \tilde{C}_{i11} (D_0 + D_{13})(m_{\tilde{b_1}}^2, m_{\tilde{b_1}}^2) \biggr] 
\nonumber \\
&& +
  \cos 2{\tilde \theta}_b \tilde{C}_{i12}
  \biggl[ (D_0 + D_{13})(m_{\tilde{b_1}}^2, m_{\tilde{b_2}}^2) + 
  (D_0 + D_{13})(m_{\tilde{b_2}}^2, m_{\tilde{b_1}}^2) \biggr] \biggr\}
\nonumber\\
X_{B_1}^{(t)} &=& -\frac{3}{2} m_{\tilde{g}} 
  \biggl\{ \sin 2{\tilde \theta}_b
  \biggl[ \tilde{C}_{i22} D_{13}(m_{\tilde{b_2}}^2, m_{\tilde{b_2}}^2) - 
          \tilde{C}_{i11} D_{13}(m_{\tilde{b_1}}^2, m_{\tilde{b_1}}^2) \biggr] 
\nonumber \\
&& +
  \cos 2{\tilde \theta}_b \tilde{C}_{i12}
  \biggl[ D_{13}(m_{\tilde{b_1}}^2, m_{\tilde{b_2}}^2) + 
  D_{13}(m_{\tilde{b_2}}^2, m_{\tilde{b_1}}^2) \biggr] \biggr\}
\nonumber\\
X_{B_1}^{(1)} &=& 3 m_{\tilde{g}} 
  \biggl\{ \sin 2{\tilde\theta}_b
  \biggl[ \tilde{C}_{i22} (D_{11} - D_{13})(m_{\tilde{b_2}}^2, m_{\tilde{b_2}}^2) - 
          \tilde{C}_{i11} (D_{11} - D_{13})(m_{\tilde{b_1}}^2, m_{\tilde{b_1}}^2) \biggr] 
\nonumber\\
&& +
   \cos 2{\tilde \theta}_b \tilde{C}_{i12}
  \biggl[ (D_{11} - D_{13})(m_{\tilde{b_1}}^2, m_{\tilde{b_2}}^2) + 
  (D_{11} - D_{13})(m_{\tilde{b_2}}^2, m_{\tilde{b_1}}^2) \biggr] \biggr\}\, ,
\label{fg:box1}
\end{eqnarray}
where the
full arguments of the scalar integral and tensor coefficients are 
$(p_1^2,p_2^2,p_3^2; m_1,m_2,m_3,m_4) = (0, 0, 0; m_{\tilde{b_i}},m_{\tilde{g}},m_{\tilde{g}}^2,
m_{\tilde{b_j}}^2)$.

{\bf{\underline{Diagram $B_2$:}}}

\begin{eqnarray}
X_{B_2}^{(s)} &=& \frac{1}{6} m_{\tilde{g}} 
  \biggl\{ \sin 2{\tilde \theta}_b
  \biggl[ \tilde{C}_{i22} D_{13}(m_{\tilde{b_2}}^2, m_{\tilde{b_2}}^2) - 
          \tilde{C}_{i11} D_{13}(m_{\tilde{b_2}}^2, m_{\tilde{b_2}}^2) \biggr] 
\nonumber\\
&& +
   \cos 2{\tilde \theta}_b \tilde{C}_{i12}
  \biggl[ D_{13}(m_{\tilde{b_1}}^2, m_{\tilde{b_2}}^2) + 
  D_{13}(m_{\tilde{b_2}}^2, m_{\tilde{b_1}}^2) \biggr] \biggr\}
 \nonumber \\
\nonumber\\
X_{B_2}^{(t)} &=& - \frac{1}{6} m_{\tilde{g}} 
  \biggl\{ \sin 2{\tilde \theta}_b
  \biggl[ \tilde{C}_{i22} D_{13}(m_{\tilde{b_2}}^2, m_{\tilde{b_2}}^2) + 
          \tilde{C}_{i11} D_{13}(m_{\tilde{b_1}}^2, m_{\tilde{b_1}}^2) \biggr] 
\nonumber\\
&& +
   \cos 2{\tilde \theta}_b \tilde{C}_{i12}
  \biggl[ D_{13}(m_{\tilde{b_1}}^2, m_{\tilde{b_2}}^2) + 
  D_{13}(m_{\tilde{b_2}}^2, m_{\tilde{b_1}}^2) \biggr] \biggr\}
\nonumber\\
X_{B_2}^{(1)} &=& \frac{1}{3} m_{\tilde{g}} 
  \biggl\{ \sin 2{\tilde\theta}_b
  \biggl[ \tilde{C}_{i22} (D_{12} - D_{13})(m_{\tilde{b_2}}^2, m_{\tilde{b_2}}^2) -
          \tilde{C}_{i11} (D_{12} - D_{13})(m_{\tilde{b_1}}^2, m_{\tilde{b_1}}^2) \biggr] 
\nonumber\\
&& +
  \cos 2{\tilde \theta}_b \tilde{C}_{i12}
  \biggl[ (D_{12} - D_{13})(m_{\tilde{b_1}}^2, m_{\tilde{b_2}}^2) + 
  (D_{12} - D_{13})(m_{\tilde{b_2}}^2, m_{\tilde{b_1}}^2) \biggr] \biggr\}\, ,
\label{fg:box2}
\end{eqnarray}
where the arguments of the scalar integral and tensor coefficients are 
$(p_1^2,p_2^2,p_3^2; m_1^2,m_2^2,m_3^2,m_4^2) = 
(0, 0, 0; m_{\tilde{b_i}}, m_{\tilde{b_i}}, m_{\tilde{g}},m_{\tilde{b_j}})$. 

\newpage
{\bf{\underline{Diagram $B_3$:}}}

\begin{eqnarray}
X_{B_3}^{(s)} &=& -\frac{1}{6} m_{\tilde{g}} 
  \biggl\{ \sin 2{\tilde \theta}_b
  \biggl[ \tilde{C}_{i22} D_{13}(m_{\tilde{b_2}}^2, m_{\tilde{b_2}}^2) - 
          \tilde{C}_{i11} D_{13}(m_{\tilde{b_2}}^2, m_{\tilde{b_2}}^2) \biggr] 
\nonumber\\
&& +
   \cos 2{\tilde \theta}_b \tilde{C}_{i12}
  \biggl[ D_{13}(m_{\tilde{b_1}}^2, m_{\tilde{b_2}}^2) + 
  D_{13}(m_{\tilde{b_2}}^2, m_{\tilde{b_1}}^2) \biggr] \biggr\}
\nonumber\\
X_{B_3}^{(t)} &=& \frac{1}{6} m_{\tilde{g}} 
  \biggl\{ \sin 2{\tilde \theta}_b
  \biggl[ \tilde{C}_{i22} D_{13}(m_{\tilde{b_2}}^2, m_{\tilde{b_2}}^2) + 
          \tilde{C}_{i11} D_{13}(m_{\tilde{b_1}}^2, m_{\tilde{b_1}}^2) \biggr] 
\nonumber\\
&& +
   \cos 2{\tilde \theta}_b \tilde{C}_{i12}
  \biggl[ D_{13}(m_{\tilde{b_1}}^2, m_{\tilde{b_2}}^2) + 
  D_{13}(m_{\tilde{b_2}}^2, m_{\tilde{b_1}}^2) \biggr] \biggr\}
\nonumber\\
X_{B_3}^{(1)} &=& \frac{1}{3} m_{\tilde{g}} 
  \biggl\{ \sin 2{\tilde \theta}_b
  \biggl[ \tilde{C}_{i22} (D_{12} - D_{13})(m_{\tilde{b_2}}^2, m_{\tilde{b_2}}^2) -
          \tilde{C}_{i11} (D_{12} - D_{13})(m_{\tilde{b_1}}^2, m_{\tilde{b_1}}^2) \biggr] 
\nonumber\\
&& +
  \cos 2{\tilde \theta}_b \tilde{C}_{i12}
  \biggl[ (D_{12} - D_{13})(m_{\tilde{b_1}}^2, m_{\tilde{b_2}}^2) + 
  (D_{12} - D_{13})(m_{\tilde{b_2}}^2, m_{\tilde{b_1}}^2) \biggr] \biggr\}\, ,
\label{fg:box3}
\end{eqnarray}
where the arguments of the scalar integral and tensor coefficients are 
$(p_1^2,p_2^2,p_3^2; m_1,m_2,m_3,m_4) = 
(0, 0, 0; m_{\tilde{b_i}}, m_{\tilde{b_i}}, m_{\tilde{g}},m_{\tilde{b_j}})$.
\bibliography{bghb_paper}
\bibliographystyle{unsrt}
\end{document}